\documentclass[aps,twocolumn,showpacs,preprintnumbers,amsmath,amssymb]{revtex4}
\usepackage{latexsym}
\usepackage{psfig,graphicx}
\usepackage{bm}
\renewcommand{\theequation}{\arabic{section}.\arabic{equation}}
\newcommand{\bea}{\begin{eqnarray}}
\newcommand{\eea}{\end{eqnarray}}
\newcommand{\be}{\begin{equation}}
\newcommand{\ee}{\end{equation}}
\newcommand{\pkt}{\; .}

\newcommand{\eqn}[1]{(\ref{#1})}
\newcommand{\tr}{{\rm Tr}}
\newcommand{\im}{{\rm Im}}

\newcommand{\calm}{{\cal M}}

\newcommand{\calv}{{\cal V}}

\newcommand{\bfp}{{\bf p }}
\newcommand{\tcsplit}{\nonumber \\   &&}

\newcommand{\call}{{\cal L}}

\newcommand*{\del}{\partial}
\renewcommand{\Re}{\textrm{Re}\,}

\renewcommand{\vec}[1]{\ensuremath{\mathbf{#1}}}

\begin{document}
\preprint{DO-TH-03/18}
\date{\today}
\title{\bf Out-of-equilibrium evolution of quantum fields in the hybrid model
with quantum back reaction}   
\author{J\"urgen Baacke}
\email{baacke@physik.uni-dortmund.de}
\author{Andreas Heinen}
\email{andreas.heinen@uni-dortmund.de}
\affiliation{Institut f\"ur Physik, Universit\"at Dortmund,
D - 44221 Dortmund , Germany}
\begin{abstract}
The hybrid model with a scalar "inflaton" field coupled to a "Higgs"
field with a broken symmetry potential is one of the promising models
for inflation and (p)reheating after inflation.
We consider the nonequilibrium evolution of the quantum fields
of this model with quantum back reaction in the Hartree approximation,
in particular the transition of the Higgs field from the metastable
"false vacuum" to the broken symmetry phase.
We have performed the renormalization of the equations of motion,
of the gap equations and of the energy density, using dimensional
regularization. We study the influence of the back reaction on the
evolution of the classical fields and of the quantum fluctuations.
We observe that back reaction plays an important role over a wide range 
of parameters. Some implications of our investigation for the 
preheating stage after cosmic inflation are presented.
\end{abstract}
\pacs{98.80.Cq, 11.30.Qc}
\maketitle

\section{\label{sec:intro}Introduction}
\setcounter{equation}{0}
The hybrid model has been proposed by Linde  
\cite{Linde:1990gz,Linde:1991km,Linde:1994cn} as a possible inflationary
scenario, several variants of the model have been discussed 
recently \cite{Copeland:1994vg,Garcia-Bellido:1997wm,Micha:1999wv,
Buchmuller:2000zm,Nilles:2001fg,Asaka:2001ez,Cormier:2001iw}.
Here we consider the model in the context of preheating after inflation,
and not inflation itself. This period has many interesting aspects
and has received a wide attention
\cite{Garcia-Bellido:1997wm,Lyth:1998xn,Garcia-Bellido:1999sv,
Bastero-Gil:1999fz,Krauss:1999ng,Felder:2000hj,Micha:1999wv,Felder:2001kt,
Copeland:2001qw,Buchmuller:2000zm,Nilles:2001fg,Asaka:2001ez,Cormier:2001iw,Garcia-Bellido:2002aj,Borsanyi:2002tm:Borsanyi:2003ib}.
 For this period it is a reasonable approximation
to neglect the coupling to gravity and the expansion of the universe.
In hybrid inflation the inflaton transfers at the end of the slow roll 
period its energy to a Higgs field, whose classical
potential changes from a symmetric potential to a double well
potential, thereby going through a period of abundant
particle production. In the  hybrid model two principal mechanisms
for particle production are at work:  spinodal instability and 
parametric resonance. Their relative importance depends on the
parameters of the particular scenario.
The hybrid  model may arise naturally in the context of
 supersymmetry and supergravity \cite{Lyth:1998xn,Bastero-Gil:1999fz,
Micha:1999wv,Buchmuller:2000zm,Asaka:2001ez};
it could also display essential features of an electroweak reheating 
\cite{Garcia-Bellido:1999sv,Garcia-Bellido:2002aj}.
Another aspect of the type of model investigated here may be that
it simulates a first order phase transition, where the change in the
effective potential does not arise from a decrease in temperature but
is mediated by an effective field. It thereby replaces models
\cite{Boyanovsky:1997rw,Bowick:1998kd,Felder:2000hj,Felder:2001kt} 
where a rapid decrease of  the temperature is simulated by an 
instantaneous quench.

In this paper we address several new topics:\newline
(i) we include the back reaction of the produced Higgs and inflaton
 quanta onto the classical Higgs and inflaton fields {\em and onto themselves 
in the Hartree approximation}.
This feature limits the exponential growth of fluctuations due to
negative squared masses and to parametric resonance, 
which is unhibited if this back reaction is not included. 
In previous investigations quantum fluctuations have been
included without back reaction \cite{Micha:1999wv},
with back reaction in the one-loop approximation
\cite{Cormier:2001iw} and in the Hartree approximation
with an UV cutoff in a 
supersymmetric model including axions \cite{Bastero-Gil:1999fz}. 
The model investigated there is rather special, it incorporates
several scalar degrees of freedom of a non-minimal
supersymmetric standard model. So on the one hand 
it has more scalar fields than
the simple hybrid model, on the other hand the couplings are 
constrained.\newline
(ii) on the formal level we present a consistent renormalization
of the gap equations arising in the Hartree approximation, making
use of the two-particle point-irreducible (2PPI) formalism 
\cite{Verschelde:1992bs:Coppens:1993zc,Baacke:2002ee,Baacke:2003qh}. 
This is a first example of
renormalization for coupled channel system out of equilibrium
 with Hartree-type self-interaction, and goes beyond the
one-loop renormalization of Ref. \cite{Cormier:2001iw} and the unrenormalized 
Hartree approximation of Ref.~\cite{Bastero-Gil:1999fz}.
\newline
(iii) we discuss, as previous authors, the approach to spontaneous
symmetry breaking; as in Ref. \cite{Garcia-Bellido:2002aj} we find that
the classical Higgs field approaches the broken symmetry minimum 
with an exponential behavior.  However, we find that for sufficiently
high excitations the system does not display a broken symmetry
phase at late times; for such excitations the broken symmetry minima 
are washed out by quantum fluctuations.

We are not able to discuss thermalization, here. Thermalization is not
supposed to occur in the Hartree approximation when the fields 
are homogeneous in space. One of the possibilities in the present context 
has been discussed in Refs. \cite{Asaka:2001ez,Garcia-Bellido:2002aj}:  
a transition between a quantum description at early times, until the 
the low energy modes reach high occupation numbers, and appending
a classical evolution of this system at late times.
For such a classical evolution in a $\Phi^4$ model in $1+1$ dimensions
thermalization has recently been reported and systematically investigated 
in  Ref.~\cite{Boyanovsky:2003tc}. In a $3+1$ dimensional
massless $\Phi^4$ model the system is found to evolve towards a turbulent
behavior at late times \cite{Micha:2002ey}. Thermalization and/or ``equilibration''
has also been discussed within a purely quantum field theoretical
description for spatially inhomogeneous fields in the Hartree approximation 
\cite{Salle:2000hd:Salle:2000jb} and for 
spatially homogeneous fields, when going to the next-to-leading order
in a $1/N$ expansion within the 2PI formalism (2PI-NLO-1/N)
\cite{Berges:2001fi,Aarts:2002dj} 
or to the bare vertex \cite{Cooper:2002qd} approximations.
The issue is not closed at present, and the different approaches
have to be studied in conjunction in the future.

The plan of the paper is as follows. 
We start by briefly reviewing the hybrid model 
 and its application to the preheating stage 
after cosmic inflation in Sec.~\ref{sec:model}. 
In Sec.~\ref{sec:effective-action} 
we present an effective action and the derivation of the equations of motion
for the nonequilibrium quantum dynamics in the hybrid model. 
In Sec.~\ref{sec:approx} we present 
approximations to the effective action and specify 
the Hartree approximation by giving the expressions for the fluctuation
integrals and the energy contributions.
The renormalization is discussed in Sec.~\ref{sec:renorm}. 
In Sec.~\ref{sec:numer-impl} we give details of the numerical implementation.
The numerical results for different parameter sets are presented and 
discussed in Sec.~\ref{sec:results}.
We end with conclusions in Sec.~\ref{sec:conclusions}. 
The paper is completed by three appendixes.


\section{The Hybrid Model}
\label{sec:model}
\setcounter{equation}{0}
\subsection{General form of the potential}
\label{sec:gener-form-potent}
\begin{figure}[htbp]
  \centering
  \includegraphics[width=8.5cm]{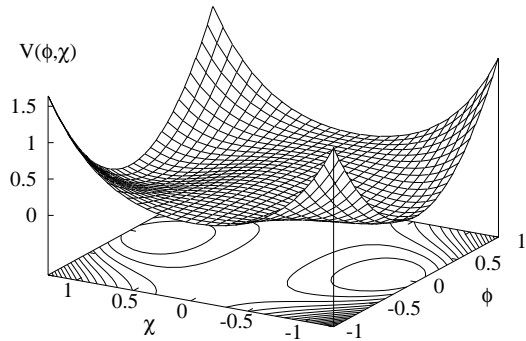}
  \caption{Generic shape of a hybrid potential $V(\phi,\chi)$ 
as a function of the classical fields $\phi$ and $\chi$; 
at the base of this plot the contour lines of the potential are show.}
  \label{fig:hybrid-potential}
\end{figure}

We consider the hybrid model as proposed by Linde \cite{Linde:1994cn} 
defined by the Lagrange density
\bea
\mathcal{L}=\frac{1}{2}\del_\mu\Phi\del^\mu \Phi+\frac{1}{2}\del_\mu
X\del^\mu X-V(\Phi,X) \ ,
\eea
with the potential (hybrid potential)
\bea
V(\Phi,X)=\frac{1}{2}m^2\Phi^2+\frac{1}{2}g^2\Phi^2 X^2
+\frac{\lambda}{4}(X^2-v^2)^2\label{eq:hybrid-pot} \ .
\eea
We will refer to $\Phi$ as the inflaton and to $X$ as the Higgs 
 or the symmetry breaking field.  
For simplicity the dependence of all fields on the space-time arguments 
$x=(t,\vec{x})$ is suppressed. The discussion will be restricted to 
a space-time with Minkowski metric. The Higgs field has a double-well 
potential with a classical vacuum expectation value given by $v$. 
Both fields $\Phi$ and $X$ are assumed to have non-vanishing 
classical expectation values (order parameters of the fields), i.e.
\bea
\left\langle \Phi\right\rangle&=&\phi \ ,\\
\left\langle X\right\rangle&=&\chi  \ .
\eea
The generic shape of the classical hybrid potential $V(\phi,\chi)$ 
is shown in Fig.~\ref{fig:hybrid-potential}. 
Classically there are the two degenerate minima at the points with 
$\phi=0$ and $\chi=\pm\lambda v$. They can easily be identified 
in the contour lines at the base of the plot 
in Fig.~\ref{fig:hybrid-potential}.
However, if the corrections from quantum fluctuations around the 
classical solutions are included,  
the minima of the interacting theory are not equal to the 
ones in the classical theory.

The Higgs field has an effective mass square 
$m^2_X(\Phi)=-\lambda v^2 + g^2 \Phi^2$. Thus the field $X$, 
i.e. the classical part $\chi$ and its quantum fluctuations, 
becomes unstable if the absolute value of the inflaton field $\Phi$ 
is lower than 
\bea
\Phi_\mathrm{c}=\frac{v\sqrt{\lambda}}{g}\label{eq:phi_c} \ .
\eea
The region with $-\Phi_\mathrm{c}<\Phi<\Phi_\mathrm{c}$ in the potential 
is called the spinodal region.

\subsection{Energy scales and parameters}
\label{sec:energy-scal-param}
If the hybrid model is taken as an effective model for the (p)reheating 
stage at the end of cosmic inflation, 
the mass parameters $m^2$ and $v^2$ and the coupling constants 
$g$ and $\lambda$ are constrained by the observations of the 
various cosmic microwave background experiments 
(see, e.g., Refs.~\cite{Kinney:2003uw,Peiris:2003ff,Barger:2003ym,Leach:2003us}
for recent analyzes). 
In order to fulfill the slow-roll condition the inflaton mass $m$ has to be 
chosen very small \cite{Linde:1994cn}
\be
m^2 \ll g^2 v^2
\pkt
\ee
This means that the potential in the $\phi$ direction 
close to $\chi\approx 0$ is very flat. Indeed when treating the period
after inflation the inflaton mass $m^2$ can be neglected
(see, e.g., \cite{Copeland:2002ku}); we have
chosen $m^2=0$ in the numerical simulations throughout. 
As discussed in \cite{Linde:1994cn}
the couplings $\lambda$ and $g^2$ can vary over a wide range, depending on
the specific preheating scenario. This is interrelated with the
choice of the mass scale $v$ which is chosen to be in the
numerical simulations. As we do not couple to gravity its absolute
physical value is irrelevant here. Of course this choice
enters into the time, momentum and energy scales.

As a concrete phenomenological model it can be embedded at least at two 
manifestly different energy scales.
If one has electroweak preheating \cite{Garcia-Bellido:1997wm,
Garcia-Bellido:1999sv,Garcia-Bellido:2002aj,Krauss:1999ng,Copeland:2001qw}
in mind, the symmetry breaking field 
mimics the standard Higgs sector of the standard model, while ignoring 
contributions from the gauge and fermion fields. 
Of course we would have to generalize $X$ to be a
complex doublet in order to represent the SU(2) symmetry breaking Higgs sector. 
The vacuum expectation value $v$ in this case would be chosen equal to 246GeV. 
The classical Higgs mass is $m_\chi^2=\lambda v^2$. 
If the phase transition at the end of inflation takes place at the scale
of a Grand Unified Theory (GUT), the symmetry breaking field $X$ 
could be, e.g., a very heavy sneutrino, i.e. 
the scalar super-partner of one of the light neutrinos. 
The reheating in such a scenario would come 
from the decay of this heavy sneutrino. 
We will not restrict our study to one of these two scenarios in the following.

\section{Effective action}
\label{sec:effective-action}
\subsection{Quantum fluctuations and their non-perturbative resummation}
\label{sec:quant-fluct}
The time evolution of out-of-equilibrium systems in quantum field theory
is described within the closed-time-path (CTP) or Schwinger-Keldysh 
formalism \cite{Schwinger:1961qe,Keldysh:1964ud}. Nonequilibrium quantum 
field theory has
seen a tremendous progress, after seminal publications in the
1980's \cite{Cooper:1986wv:Eboli:1988fm}, 
mainly in the last decade
\cite{Boyanovsky:1993pf:Boyanovsky:1998zg,Boyanovsky:1997rw,Boyanovsky:1995me,
Cooper:1994hr:Cooper:1997ii,Baacke:1997se,Baacke:1998di,Baacke:2000fw}.
The main ingredient, besides the CTP formalism itself, are various
approximations based on variational principles. The one-loop,
Hartree and and large-$N$ approximations can be based
on the two-particle irreducible (2PI) formalism 
\cite{Luttinger:1960ua:deDominicis:1964,Cornwall:1974vz}. 
This formalism has also been used
beyond these mean field approaches \cite{Berges:2000ur,Berges:2001fi,
Berges:2002cz,Cooper:2002qd,Ikeda:2004in}. As an alternative
approach beyond leading orders we have recently adapted 
\cite{Baacke:2002ee,Baacke:2003qh} 
the so-called two-particle point irreducible (2PPI)
 formalism \cite{Verschelde:1992bs:Coppens:1993zc,Verschelde:2000dz}
to nonequilibrium quantum field theory.

Within the framework outlined above there are several ways for taking
into account  the quantum fluctuations and their 
\emph{back reaction} onto the classical (mean) fields and onto themselves. 
In any case they should  to be  treated non-perturbatively, 
as the amplitudes of the fields can become very large. 
In practice one resums an infinite number of perturbative Feynman diagrams via 
a Schwinger-Dyson equation. 
In this section we will derive all equations of motion, i.e. the classical 
and quantum equations, from an effective action that incorporates 
full back reaction of the quantum modes.
 
Within nonequilibrium quantum field theory the existence of a 
conserved energy is automatically guaranteed if all equations of motion 
are derived from the same variational functional, i.e. the effective action.
In the context of relativistic quantum field theory Cornwall, Jackiw 
and Tomboulis (CJT) have derived the 2PI effective action in 
Ref. \cite{Cornwall:1974vz}. In addition to a local source term 
$J(x)$ for a \emph{mean field} $\phi(x)$ one introduces a bilocal source term 
$K(x,x')$ for the Operator 
$\left<\Phi(x)\Phi(x')\right>=G(x,x')+\phi(x)\phi(x')$ 
in the generating functional. 
The effective action $\Gamma$ follows from a Legendre transformation of the 
generating functional with variational parameters $\phi(x)$ and $G(x,x')$. 
The leading orders in the 2PI or CJT action 
exactly reproduce the well know one-loop, large-$N$ and Hartree 
approximations.  

A related effective action is the 2PPI 
effective action \cite{Verschelde:1992bs:Coppens:1993zc,Verschelde:2000dz}. 
There one has a local source term $K(x,x')=K(x)\delta(x-x')$ 
for the composite operator 
$\left<\Phi^2(x)\right>$ thus making all self-energy insertions into the 
propagator $G(x,x')$ local.
In the 2PPI effective action formalism the renormalization of the 
Hartree (and leading order large-$N$) approximation is very transparent, 
because the non-perturbative counterterms in the 
effective theory are mapped one-to-one to 
the standard counterterms of a perturbative expansion of the basic Lagrangian.
In addition the 2PPI scheme allows a systematic improvement of the Hartree 
approximation without all the computational complications from 
nonlocal self-energies as introduced beyond leading oder in the 2PI scheme.

\subsection{The two-particle point-irreducible formalism}
\label{sec:two-particle-point}
At first, a slightly more general Lagrange density for a $N$ component field 
$\mathbf{\Phi}=(\Phi_1,\ldots,\Phi_N)$ is given by
\bea
\mathcal{L}(x)&=&\frac{1}{2}\del_\mu \Phi^i(x)\del^\mu\Phi^i(x)
-V(x)\label{eq:general-lag}
\ ,\\
V(x)&=&\frac{1}{2}m_{ij}^2\Phi^i(x)\Phi^j(x)
\tcsplit
+\frac{1}{4!}\lambda_{ijkl}\Phi^i(x)\Phi^j(x)\Phi^k(x)\Phi^l(x)
\label{eq:general-pot}
\ ,
\eea
where a summation over $i,j,k,l=1,\ldots,N$ etc. is understood 
and we have introduced a coupling constant matrix $\lambda_{ijkl}$ 
and a mass matrix $m_{ij}^2$. The fields $\Phi^i$ may have classical 
expectation values, i.e. $\left<\Phi^i\right>=\phi^i$.

The standard hybrid potential in Eq.~(\ref{eq:hybrid-pot}) follows from
Eq.~(\ref{eq:general-pot}) for the case $N=2$ with the identifications
\begin{eqnarray}
\Phi_1&=&\Phi\label{eq:id1}\\
\Phi_2&=&X\label{eq:id2}\\
m_{11}^2&=&m^2\label{eq:id3}\\
m_{22}^2&=&-\lambda v^2\label{eq:id4}\\
m_{12}^2&=&m_{21}^2=0\label{eq:id5}\\
\lambda_{2222}&=&6\lambda\label{eq:id6}\\
\lambda_{\mathrm{Perm}(1122)}&=&2g^2\label{eq:id7}\\
\lambda_{1111}&=&\lambda_{1112}=\ldots =\lambda_{2221}=0 \label{eq:id8}\ .
\end{eqnarray}
The 2PPI formalism is formulated in terms of the mean fields $\phi^i$ and 
local propagator insertions $\Delta_{ij}$.
The local insertions in the propagator are resummed via a Schwinger-Dyson 
or gap equation.
The Green's function $G$ fulfills the local equation
\be
(G^{-1})_{ij}(x,x')=i\left(\square \delta_{ij}+\calm^2_{ij}(x)\right) 
\delta^{(D)}(x-x') \ , \ 
\ee
where $\calm^2_{ij}$ is a variational mass parameter explained below.

For the general potential in Eq.~(\ref{eq:general-pot}) 
the 2PPI effective action can be written as 
(see e.g. Ref.~\cite{Verschelde:2000ta,Verschelde:2000dz})
\bea
\Gamma[\phi^i,\Delta^{ij}]&=&S[\phi^i]
+\Gamma^\mathrm{2PPI}[\phi^i,\calm^2_{ij}]
\tcsplit
+\frac{1}{8}\lambda_{ijkl}\int {d^Dx}\Delta^{ij}(x)\Delta^{kl}(x) \label{eq:Gamma}
\eea
with
\bea
\Delta^{ij}(x)&=&-2\frac{\delta \Gamma^\mathrm{2PPI}}{\delta
  \calm^2_{ij}(x)} \ .
\eea
The masses $\calm^2_{ij}(x)$ have to fulfill the so called \emph{gap equations}
\bea
\calm^2_{ij}(x)&=&m_{ij}^2+\frac{1}{2}\lambda_{ijkl}
\left(\phi^k(x)\phi^l(x) +\Delta^{kl}(x)\right) \ . \
\eea
The term $\Gamma^\mathrm{2PPI}[\phi^i,\calm^2_{ij}]$ denotes 
the infinite sum over all 2PPI diagrams. A diagram that does not fall apart if 
two lines meeting at the same point are cut is called 
\emph{two-particle point-irreducible}.
  
The classical action $S[\phi^i]$ is given by
\bea
S[\phi^i]&=&\int d^D x \left[\frac{1}{2}\del_\mu \phi^i(x)\del^\mu \phi^i(x)-
\frac{1}{2}m_{ij}^2\phi^i(x)\phi^j(x)\right.\nonumber \\
&&\qquad+\left.\frac{1}{4!}\lambda_{ijkl}\phi^i(x)\phi^j(x)\phi^k(x)\phi^l(x)\right] \ .
\eea
The term proportional to an integral over 
$\Delta^{ij}(x)\Delta^{kl}(x)$ in Eq.~(\ref{eq:Gamma}) 
corrects the double counting of bubble graphs.

With the help of the identifications in Eq.~(\ref{eq:id1})--(\ref{eq:id8}) 
the 2PPI effective action for the hybrid potential in Eq.~(\ref{eq:hybrid-pot})reads explicitly
\bea
&&\hspace{-1cm}\Gamma[\phi,\chi,\Delta_{\phi\phi}
,\Delta_{\phi\chi},\Delta_{\chi\chi}]
\tcsplit
=S[\phi,\chi]
\tcsplit
+\Gamma^\mathrm{2PPI}\left[\phi,\chi,\calm^2_{\phi\phi}
,\calm^2_{\phi\chi},\calm^2_{\chi\chi}\right]\nonumber \\
&&+\frac{3 \lambda}{4} \int d^Dx \Delta_{\chi\chi}^2(x)
\tcsplit
+\frac{g^2}{2}\int
d^Dx\left(\Delta_{\phi\phi}(x)\Delta_{\chi\chi}(x)+2\Delta_{\phi\chi}^2(x)\right)\ .
\eea
The gap equations for the masses $\calm^2$ are
\bea
\calm^2_{\phi\phi}(x)&=&m^2+g^2\left(\chi^2(x)+\Delta_{\chi\chi}(x)\right)
\label{eq:gap-equation1}\ ,\\
\calm^2_{\chi\chi}(x)&=&-\lambda v^2+3\lambda
\left(\chi^2(x)+\Delta_{\chi\chi}(x)\right)\nonumber \\
&&\qquad +g^2\left(\phi^2(x)+\Delta_{\phi\phi}(x)\right)\label{eq:gap-equation2}\ ,\\
\calm^2_{\phi\chi}(x)&=&\calm^2_{\chi\phi}(x)
=2g^2\left(\phi(x)\chi(x)+\Delta_{\phi\chi}(x)\right)\ 
\ .\ \label{eq:gap-equation3}
\eea
The self-energy insertions $\Delta$ are 
\bea
\Delta_{\phi\phi}(x)&=&-2\frac{\delta \Gamma^\mathrm{2PPI}}{\delta
  \calm^2_{\phi\phi}(x)} \ ,\\
\Delta_{\chi\chi}(x)&=&-2\frac{\delta \Gamma^\mathrm{2PPI}}{\delta
  \calm^2_{\chi\chi}(x)} \ ,\\
\Delta_{\phi\chi}(x)&=&-2\frac{\delta \Gamma^\mathrm{2PPI}}{\delta
  \calm^2_{\phi\chi}(x)} \ .
\eea
Inverting the gap equations gives
\bea
\Delta_{\phi\phi}(x)&=&\frac{1}{g^2}\left(\calm^2_{\chi\chi}(x)+\lambda
  v^2-\frac{3\lambda}{g^2}\left(\calm^2_{\phi\phi}(x)-m^2\right)
\right)\nonumber \\
&&-\phi^2(x) \ ,\\
\Delta_{\chi\chi}(x)&=&\frac{1}{g^2}
\left(\calm^2_{\phi\phi}(x)-m^2\right)-\chi^2(x)\ ,\\
\Delta_{\phi\chi}(x)&=&\frac{1}{2g^2}\calm^2_{\phi\chi}(x)-\phi(x)\chi(x) \ .
\eea
The effective action can therefore be expressed in a form without the
$\Delta$'s.
\bea
&&\hspace{-1cm}\Gamma\left[\phi,\chi,\calm^2_{\phi\phi}
,\calm^2_{\chi\chi},\calm^2_{\phi\chi}\right]\nonumber \\
&=&\int
d^Dx \bigg[\frac{1}{2}\del_\mu \phi(x)\del^\mu
\phi(x)+\frac{1}{2}\del_\mu \chi(x)\del^\mu \chi(x)\nonumber \\
&&-\frac{1}{2}\calm^2_{\phi\phi}(x)\phi^2(x)+g^2\phi^2(x)\chi^2(x)
\nonumber \\
&&-\frac{1}{2}\calm^2_{\chi\chi}(x)\chi^2(x)
+\frac{\lambda}{2}\chi^4(x)-\frac{\lambda}{4}v^4\nonumber\\
&&-\calm^2_{\phi\chi}(x)\phi(x)\chi(x)
\tcsplit
-\frac{3\lambda}{4g^4}\left(\calm^2_{\phi\phi}(x)-m^2\right)^2
\nonumber \\
&&+\frac{1}{2g^2}\left(\calm^2_{\phi\phi}(x)-m^2\right)
\left(\calm^2_{\chi\chi}(x)+\lambda
v^2\right)
\tcsplit
+\frac{1}{4g^2}\left(\calm^2_{\phi\chi}(x)\right)^2\bigg]\nonumber \\
&&+\Gamma^\mathrm{2PPI}
\left[\phi,\chi,\calm^2_{\phi\phi},\calm^2_{\phi\chi}
,\calm^2_{\chi\chi}\right] \label{eq:Gamma-2PPI}
\eea
The latter form of the effective action makes it easy 
to derive a renormalized effective action.  

\subsection{Equations of motion}
\label{sec:equations-motion}
The equations of motion for the classical fields $\phi$ and $\chi$ follow from
\bea
\frac{\delta \Gamma }{\delta \phi(x)}=0\quad \textrm{and} 
\quad \frac{\delta \Gamma }{\delta \chi(x)}=0\ . \\
\eea
If we take only the causal part of all functions by restricting them 
to the positive time branch of a closed time path 
(CTP; for the Schwinger-Keldysh closed time path formalism see 
Ref.~\cite{Schwinger:1961qe,Keldysh:1964ud,Jordan:1986ug,Calzetta:1988cq}),
we get the following classical equations of motion 
\bea
0&=&\square\phi+\calm^2_{\phi\phi}\phi
+\calm^2_{\phi\chi}\chi-2g^2\chi^2\phi
\tcsplit
-\frac{\delta
  \Gamma^\mathrm{2PPI}}{\delta \phi} \ ,\\
0&=&\square
\chi+\calm^2_{\chi\chi}\chi+\calm^2_{\phi\chi}\phi
-2\lambda \chi^3-2g^2\phi^2\chi
\tcsplit
-\frac{\delta
  \Gamma^\mathrm{2PPI}}{\delta \chi} \ .
\eea
The quantum equations of motion, i.e. the Schwinger-Dyson equations or  
 gap equations, are derived from
\bea
\frac{\delta \Gamma }{\delta \calm^2_{\phi\phi}(x)}=0\,,\, 
\frac{\delta \Gamma }{\delta \calm^2_{\chi\chi}(x)}=0\ \textrm{and}\  
\frac{\delta \Gamma }{\delta \calm^2_{\phi\chi}(x)}=0 \ . 
\eea
They have already been denoted in 
Eq.~(\ref{eq:gap-equation1})--(\ref{eq:gap-equation3}). In
general the gap equations form a coupled system of self-consistent 
nonlinear equations. Due to the self-consistency an infinite number of 
one-particle irreducible (1PI) diagrams is resummed.

In the following we will assume spatially homogeneous fields. 
The classical fields obey $\phi=\phi(t)$ and $\chi=\chi(t)$ 
and the Green's functions can be expressed via their Fourier components
\bea
G_{ij}(t,t';\vec{x},\vec{x}')=\int\frac{d^{D-1}p}{(2\pi)^{D-1}} 
e^{i\vec{p}\cdot(\vec{x}-\vec{x}')}G_{ij}(t,t';\vec{p}) \ .
\eea
Since the 2PPI formalism resums \emph{local} self energy insertions 
in the Green's function, $G_{ij}(t,t';\vec{p})$  can be rewritten in 
terms of mode functions
$f_i(t;\vec{p})$ and $f_j(t';\vec{p})$. If we do so, 
we have a Wronskian matrix, that is diagonal. 
The quantization of the theory at $t=0$ is explained in more detail 
in Sec.~\ref{subsec:initial}. The Green's function reads
\bea
G_{ij}(t,t';\vec{p})&=&\sum_{\alpha=1}^2
\frac{1}{2\omega_\alpha}\bigg[ 
f_{i}^\alpha(t;p)f_{j}^{*\,\alpha}(t';p)\Theta(t-t')
\nonumber \\
&&
+f_{i}^\alpha(t';p)f_{j}^{*\,\alpha}(t;p)\Theta(t'-t)\bigg]
\label{eq:G-factorized} \ ,
\eea
where the mode functions satisfy 
\bea
\ddot{f}_i^\alpha(t;p)+\left[\vec{p}^2\delta_{ij}+\calm^2_{ij}(t)
\right]f_j^\alpha(t;p)=0 \label{eq:mode-equations}\ .
\eea
The fundamental solutions of this system of coupled differential 
equations will be labeled with Greek letters $\alpha,\beta,\ldots$. 
So there are four different complex mode functions.

In Eq.~(\ref{eq:G-factorized}) we have introduced the quantities 
$\omega_\alpha$ defined by
\bea
\omega_\alpha=\sqrt{\vec{p}^2+m_{0,\,\alpha}^2} \ .
\eea
They will be explained below, as they depend on the initial conditions
via the initial masses $m^2_{0,\,\alpha}$.

The explicit form of the equations of motion depends on a given 
approximation for $\Gamma^\mathrm{2PPI}$. 
In addition one has to renormalize the equations of motion for all
approximation schemes beyond the classical, i.e. the tree level.


\section{\label{sec:approx}Approximations}
\setcounter{equation}{0}
The infinite sum of two-particle point-irreducible diagrams in 
$\Gamma^\mathrm{2PPI}$ can be truncated in several ways. 
We will discuss a loop expansion here. 
The loop expansion can be formulated as
\bea
\Gamma^\mathrm{2PPI}=\sum_{\ell=1}^\infty\Gamma^{(\ell)}
=\Gamma^{(1)}+\Gamma^{(2)}+\ldots \ ,
\eea
where the index $\ell=1,2,\ldots$ corresponds to the number of loops 
in a 2PPI diagram. The dots in the last equation indicates 
all contributions beyond the two-loop order. 
The fluctuation integrals $\Delta_{ij}$ are expanded analogous as
\bea
\Delta_{ij}(t)=\Delta_{ij}^{(1)}(t)+\Delta_{ij}^{(2)}(t)+\ldots. 
\eea
The one-loop order in the 2PPI loop expansion is equivalent 
to the Hartree approximation, as already stated above. 

\paragraph{Zero-loop -- classical approximation.}
To zero-loop order one discards the term $\Gamma^\mathrm{2PPI}$ 
completely, which leads to $\Delta_{ij}=0$ and 
$\frac{\delta \Gamma^\mathrm{2PPI}}{\delta \phi}
=\frac{\delta \Gamma^\mathrm{2PPI}}{\delta \chi}=0$.

The zero-loop contribution to the energy denotes 
\bea
E^{(0)}(t)&=&\frac{1}{2}\dot{\phi}^2(t)
+\frac{1}{2}\calm^2_{\phi\phi}(t)\phi^2(t)
-g^2\phi^2(t)\chi^2(t)\nonumber
\\
&&+\frac{1}{2}\dot{\chi}^2(t)
+\frac{1}{2}\calm^2_{\chi\chi}(t)\chi^2(t)
-\frac{\lambda}{2}\chi^4(t)+\frac{\lambda}{4}v^4\nonumber
\\
&&+\calm^2_{\phi\chi}(t)\phi(t)\chi(t)
-\frac{1}{4g^2}\left(\calm^2_{\phi\chi}(t)\right)^2\nonumber \\
&&+\frac{3\lambda}{4g^4}\left(\calm^2_{\phi\phi}(t)-m^2\right)^2
\label{eq:E-zero}
\\
&&-\frac{1}{2g^2}\left(\calm^2_{\phi\phi}(t)-m^2\right)
\left(\calm^2_{\chi\chi}(t)+\lambda
v^2\right)\nonumber  \ .
\eea

\begin{figure}[htbp]
  \centering
  \includegraphics[scale=1.2]{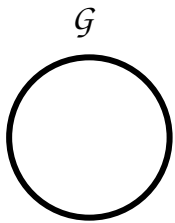}
  \caption{One-loop bubble diagram representing the Hartree approximation. 
The line denotes a propagator $\mathcal{G}$ as given by Eq.~(\ref{eq:G-matrix})}
\label{fig:bubble}
\end{figure}

\paragraph{One-loop -- Hartree approximation.}
The one-loop approximation is identical to what we will call \emph{Hartree 
approximation}. The sum of all two-particle point-irreducible 
diagrams is truncated at $\Gamma^\mathrm{2PPI}\approx \Gamma^{(1)}$ with 
(see the diagram in Fig.~\ref{fig:bubble})
\bea
\Gamma^{(1)}[\calm^2_{\phi\phi},\calm^2_{\chi\chi},\calm^2_{\phi\chi}]
=\frac{i}{2}\mathrm{Tr}\ln\left[\mathcal{G}^{-1}\right]
\ .
\eea
We have introduced
\bea
\mathcal{G}=\begin{pmatrix}
G_{\phi\phi}&G_{\phi\chi}\cr
G_{\chi\phi}&G_{\chi\chi}
\end{pmatrix} \label{eq:G-matrix}\ .
\eea

The functional derivative of $\Gamma^{(1)}$ with respect to $\calm^2_{ij}$ 
gives the tadpole insertions
\bea
\Delta_{ij}^{(1)}(t)&=&
\frac{1}{2}\int\frac{d^{D-1}p}{(2\pi)^{D-1}}
\bigg[G_{ij}(t,t;\vec{p})+G_{ji}(t,t;\vec{p})\bigg]\label{eq:Delta_ij_1}\\
&=&\sum_{\alpha=1}^2\int\frac{d^{D-1}p}{(2\pi)^{D-1}}
\frac{1}{2\omega_\alpha}\mathrm{Re}
\left[f_i^\alpha(t;p)f_j^{\alpha\,*}(t;p)\right]\nonumber \ .\\
&&
\eea
There a three different insertions, $\Delta_{\phi\phi}^{(1)}$, 
$\Delta_{\chi\chi}^{(1)}$ and 
$\Delta_{\phi\chi}^{(1)}=\Delta_{\chi\phi}^{(1)}$. The 
gap equations~(\ref{eq:gap-equation1})--(\ref{eq:gap-equation3}) resum 
one-loop bubble graphs, 
because the Green's functions $G_{ij}$ 
depend on the variational mass parameters 
$\calm^2_{ij}$. At one-loop order we have 
$\frac{\delta \Gamma^\mathrm{2PPI}}{\delta \phi}=
\frac{\delta \Gamma^\mathrm{2PPI}}{\delta \chi}=0$.

The contribution of the bubble graphs to the energy is 
defined by the relation
\bea
\frac{dE^{(1)}(t)}{dt}&=&-\frac{\delta
    \Gamma^{(1)}[\calm^2_{\phi\phi},\calm^2_{\chi\chi}
,\calm^2_{\phi\chi}]}{\delta \calm^2_{ij}(t)}
\frac{d \calm^2_{ij}(t)}{dt}\\
&=&\frac{1}{2}\int\frac{d^{D-1}p}{(2\pi)^{D-1}}G_{ij}(t,t;\vec{p})
\frac{d\calm^2_{ij}(t)}{dt}\nonumber 
\eea
This equation can be integrated explicitly if one uses the equations
of motion for the mode functions $f_i^\alpha(t;p)$, yielding
\bea
E^{(1)}(t)&=&\frac{1}{2}\int\frac{d^{D-1}p}{(2\pi)^{D-1}}
\sum_\alpha\frac{1}{2\omega_\alpha}
\tcsplit
\times
\Bigg\{\mathrm{Re}
\left[\dot{f}_\phi^\alpha(t;p)
\dot{f}_\phi^{\alpha\,*}(t;p)\right]
\tcsplit
\qquad+\mathrm{Re}\left[\dot{f}_\chi^\alpha(t;p)
\dot{f}_\chi^{\alpha\,*}(t;p)
\right]
\nonumber \\
&&\qquad
+\left(\vec{p}^2+\calm^2_{\phi\phi}(t)\right)
\mathrm{Re}\left[f_\phi^\alpha(t;p)f_\phi^{\alpha\,*}(t;p)\right]
\nonumber\\
&&\qquad
+\left(\vec{p}^2+\calm^2_{\chi\chi}(t)\right)
\mathrm{Re}\left[f_\chi^\alpha(t;p)f_\chi^{\alpha\,*}(t;p)\right]
\nonumber
\\
&&\qquad+2\calm^2_{\phi\chi}(t)\mathrm{Re}
\left[f_\phi^\alpha(t;p)f_\chi^{\alpha\,*}(t;p)\right]\Bigg\}
\label{eq:energy-E1} \ .
\eea
The momentum integrations in the quantities $\Delta^{(1)}$ and $E^{(1)}$ 
are divergent and thus have to be renormalized properly. 
We will discuss this issue in the next section.

\begin{figure}[htbp]
  \centering
  \includegraphics[scale=1.2]{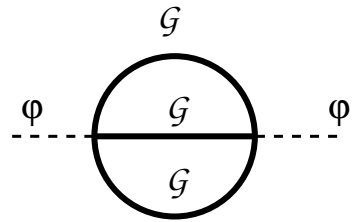}
  \caption{Two-loop sunset diagram; the solid lines denote the propagator 
$\mathcal{G}$, the dashed lines the classical 
fields $\varphi=(\phi,\chi)$.}
\label{fig:sunset}
\end{figure}

\paragraph{Two-loop -- sunset diagrams}
The only two-loop diagram appearing is the sunset diagram displayed in 
Fig.~\ref{fig:sunset}. This sunset graph leads to time integrations 
over the past of classical and quantum fields 
(``memory integrations'') and introduces scattering 
of the quanta (see Ref. \cite{Baacke:2002ee,Baacke:2003qh} 
for two-loop simulations of $\Phi^4$ theory in 1+1 dimensions). 
While it would be interesting to study how this next-to-leading 
order diagram affects the dynamics studied here, this is beyond the 
scope of this work. 

\section{\label{sec:renorm}Renormalization}
\setcounter{equation}{0}

\subsection{\label{subsec:initial}Initial conditions}
The choice of initial conditions for the quantum system 
has to be discussed together with its renormalization. 
We will take a Gaussian initial density matrix with non-vanishing 
initial values 
$\phi(t=0)$ and $\chi(t=0)$ for the classical field amplitudes.  
For the renormalization of the equations of motion 
we need a properly quantized system at the initial time. 
In order to satisfy  the usual canonical commutation relations 
for the creation and annihilation operators of the quantum fields, 
we choose a Fock space basis at $t=0$. The basic quanta
are defined by diagonalizing the mass matrix at $t=0$ and by choosing
canonical initial conditions (see below) for the mode
functions (see e.g. \cite{Baacke:1997se,Baacke:2000fw,
Cormier:2001iw,Nilles:2001fg})

We define the initial masses $m_{0,\alpha}$ as the eigenvalues
of the initial mass matrix $\calm_{ij}(0)$, i.e.,
by the equation
\bea
m_{0,\,\alpha}^2
f_i^\alpha(0;p)-\calm^2_{ij}(0)f_j^\alpha(0;p)&=&0 \ .
\eea
The eigenvalues are given by
\bea
m_{0,\,\alpha}^2&=&\frac{1}{2}\bigg[\calm^2_{\phi\phi}(0)
+\calm^2_{\chi\chi}(0)
\tcsplit
\pm \sqrt{\left(
\calm^2_{\phi\phi}(0)-\calm^2_{\chi\chi}(0)
\right)^2
+4\calm^4_{\phi\chi}(0)}\bigg] \ .
\eea
We denote the corresponding eigenvectors by $O_{i\alpha}$, where the
index $\alpha$ refers to the eigenvalue, and the latin indices to the
components. The canonical initial conditions at $t=0$ for the mode 
functions  are  
\bea
f_i^\alpha(0;p)&=&O_{i\alpha}\\
\dot{f}_i^\alpha(0;p)&=&-i\omega_\alpha
O_{i\alpha}=-i\omega_\alpha f_i^\alpha(0;p)\label{eq:modes-ini} \ .
\eea
The Wronskian  matrix of these mode functions is then
given by
\bea
W(f^\alpha_i,f^\beta_i)&=&
[f_i^{\alpha,\,*}(0;p)\dot{f}_i^\beta(0;p)
\tcsplit
-\dot{f}_i^{\alpha,\,*}(0;p)f_i^\beta(0;p) ]\\
&=&-i[(\omega_\alpha+\omega_\beta)
O_{i\alpha}O_{i\beta}] \ .
\eea
As the eigenvectors are orthogonal this matrix is diagonal.
Choosing the normalization
\be
O_{i\alpha}O_{i\beta}=\delta_{\alpha\beta}
\ee
the Wronskian matrix becomes
\bea
W(f^\alpha_i,f^\beta_i)_{\alpha,\beta=1,2}&=&-2i
\begin{pmatrix}
\omega_1&0\cr
0&\omega_2
\end{pmatrix}
\pkt \eea
If $\calm^2_{\chi\chi}(0)>\calm^2_{\phi\phi}(0)$ we can fix the matrix
$O$ as
\bea
O&=&
\begin{pmatrix}
\cos\vartheta&\sin\vartheta \cr
-\sin\vartheta&\cos\vartheta 
\end{pmatrix}\ ,\\
\tan\vartheta&=&\frac{1}{2\calm^2_{\phi\chi}(0)}
\bigg\{\calm^2_{\chi\chi}(0)-\calm^2_{\phi\phi}(0)
\tcsplit
\quad
+\sqrt{\left[\calm^2_{\chi\chi}(0)
-\calm^2_{\phi\phi}(0)\right]^2+4\calm^4_{\phi\chi}(0)
  }\bigg\}
\eea
For the opposite case $\calm^2_{\chi\chi}(0)<\calm^2_{\phi\phi}(0)$ one should
interchange $m_{0,\,1}^2$ with $m_{0,\,2}^2$ and switch 
$\vartheta\to -\vartheta$.


\subsection{\label{subsec:diverg}Isolation of the divergent contributions}
The propagator insertions $\Delta_{ij}^{(1)}$ in Eq.~(\ref{eq:Delta_ij_1}) 
are divergent. 
First we have to isolate the divergent contributions from the finite 
parts. Then the renormalization can be performed within 
a suitable regularization scheme.

A strategy for the isolation of the divergences via the perturbative 
expansion of the mode functions in terms of partial integrations 
has been described in Refs.~\cite{Baacke:1997se,Baacke:1997kj}. 
We have summarized 
what we need in Appendix~\ref{sec:pert-expans}.

In dimensional regularization with the abbreviation 
$L_\epsilon=\frac{2}{\epsilon}-\gamma+\ln 4\pi$ we have, 
using Eq.~(\ref{eq:Delta-int-div}) from  Appendix~\ref{sec:pert-expans}
and Eq.~\eqn{eq:fi-fish} and \eqn{eq:fi-tadpole} from 
Appendix~\ref{sec:feynman-iden}
\bea
\Delta_{ij}^{(1)}(t)&=&\frac{1}{2}\int\frac{d^{D-1}p}{(2\pi)^{D-1}}
\bigg[G_{ij}(t,t;\vec{p})+G_{ji}(t,t;\vec{p})\bigg]\nonumber
\\
&=&-\sum_\alpha\bigg[\frac{m_{0,\,\alpha}^2}{16\pi^2}
\bigg(L_\epsilon-\ln\frac{m_{0,\,\alpha}^2}{\mu^2}+1\bigg)
O_{i\alpha}O_{j \alpha}
\tcsplit
\quad
+\frac{1}{16\pi^2} \bigg(L_\epsilon
-\ln\frac{m_{0,\,\alpha}^2}{\mu^2}+1\bigg)
\nonumber \\
&&\quad
+\frac{1}{16\pi^2}\sum_\beta\bigg(\frac{m_{0,\,\beta}^2}{m_{0,\,\alpha}^2
-m_{0,\,\beta}^2}\ln\frac{m_{0,\,\beta}^2}{m_{0,\,\alpha}^2}\bigg)
\tcsplit
\quad\,\times\,
O_{i\alpha}O_{j\beta}O_{l\beta}O_{k\alpha}\mathcal{V}_{kl}(t)
\bigg]
\quad +\ldots\nonumber \\
&=&-\frac{1}{16\pi^2} \left[L_\epsilon+1\right]
\calm^2_{ij}(t)
\nonumber \\
&&
+\sum_\alpha\frac{m_{0,\,\alpha}^2}{16\pi^2}
O_{i\alpha}O_{j\alpha}\ln\frac{m_{0,\,\alpha}^2}{\mu^2}
\nonumber \\
&&
+\frac{1}{16\pi^2}\sum_{\alpha,\beta}
O_{i\alpha}O_{j\beta}O_{l\beta}O_{k\alpha}\mathcal{V}_{kl}(t)
\tcsplit
\times
\bigg(\ln\frac{m_{0,\,\alpha}^2}{\mu^2}
-\frac{m_{0,\,\beta}^2}{m_{0,\,\alpha}^2
-m_{0,\,\beta}^2}\ln\frac{m_{0,\,\beta}^2}{m_{0,\,\alpha}^2}\bigg)
\quad
\nonumber \\
&& +\ldots \label{eq:Delta_ij-div}\ ,
\eea
where $\sum_\alpha O_{i\alpha}O_{j\alpha}m_{0,\,\alpha}^2=\calm^2_{ij}(0)$ 
has been used. The potential $\mathcal{V}_{ij}(t)$ is defined as  
\bea
\mathcal{V}_{ij}(t)&=&\calm^2_{ij}(t)-\calm^2_{ij}(0)\ .
\eea
We can define the finite part of $\Delta_{ij}^{(1)}$ by a subtraction as 
[see Eq.~(\ref{eq:Delta-int-div})]
\bea
\Delta_{ij,\,\mathrm{fin}}^{(1)}(t)
&=&\int\frac{d^{D-1}p}{(2\pi)^{D-1}}\sum_{\alpha,\beta}
\frac{1}{2\omega_\alpha}
\tcsplit
\times
\bigg\{\mathrm{Re}
\left[f_i^\alpha(t;p)f_j^{\alpha,\,*}(t;p)\right]
\delta_{\alpha\beta}\nonumber \\
&&\quad
-O_{i\alpha}O_{j\alpha}\delta_{\alpha\beta} \label{eq:Delta_fin}\\
&&\quad +\frac{1}{\omega_\beta(\omega_\alpha+\omega_\beta)}
O_{i\alpha}O_{j\beta}O_{l\beta}O_{k\alpha}\mathcal{V}_{kl}(t)\bigg\}
\nonumber \ .
\eea
The momentum integrations in this expression are convergent, 
because the subtracted terms cancel exactly the divergent parts.
There are different finite contributions 
from the divergent part of $\Delta_{ij}^{(1)}$ 
(see. Eq.~\eqn{eq:Delta_ij-div}). 
We find it useful to define the following quantities
\bea
C^{0}_{ij}&=&\frac{1}{16\pi^2}\sum_\alpha
O_{i\alpha}O_{j\alpha}m_{0,\,\alpha}^2\ln\frac{m_{0,\,\alpha}^2}{\mu^2}\\
C^{k\ell}_{ij}&=&\frac{1}{16\pi^2}\sum_{\alpha,\beta}
O_{i\alpha}O_{j\beta}O_{k\alpha}O_{\ell\beta}
\tcsplit
\times
\bigg(\ln\frac{m_{0,\,\alpha}^2}{\mu^2}
-\frac{m_{0,\,\beta}^2}{m_{0,\,\alpha}^2
-m_{0,\,\beta}^2}\ln\frac{m_{0,\,\beta}^2}{m_{0,\,\alpha}^2}\bigg)
\eea
and 
\bea
C^1_{ij}:=C^{11}_{ij}\, ,\quad C^2_{ij}:=C^{22}_{ij}\, ,\quad C^3_{ij}:=C^{12}_{ij}+C^{21}_{ji} \ .
\eea
Thus the full 2PPI insertion $\Delta_{ij}^{(1)}$ takes 
a very simple form given by
\bea
\Delta_{ij}^{(1)}(t)&=&\Delta_{ij,\mathrm{fin}}^{(1)}(t)-\frac{1}{16\pi^2} 
\left[L_\epsilon+1\right]
\calm^2_{ij}(t)\nonumber \\
&&+C^{0}_{ij}
+C^{1}_{ij}\left(\calm^2_{\phi\phi}(t)-\calm^2_{\phi\phi}(0)\right)
\nonumber \\ 
&&
+C^{2}_{ij}\left(\calm^2_{\chi\chi}(t)-\calm^2_{\chi\chi}(0)\right)
\nonumber \\
&&+C^{3}_{ij}\left(\calm^2_{\phi\chi}(t)-\calm^2_{\phi\chi}(0)\right) 
\label{eq:Delta_ij-isol}\ .
\eea
As one can see the divergent part of $\Delta_{ij}^{(1)}$ is directly 
proportional to $\calm^2_{ij}$, i.e. with a uniform factor for all 
combinations of $i$ and $j$. In particular it is independent of 
the masses $m_{0,\,\alpha}^2$ and the matrix $O_{ij}$ and 
thereby of the initial conditions. On the other hand the different 
finite parts depend on the initial conditions via the constants $C_{ij}^{n}$ 
and involve all components of the effective mass matrix $\calm^2_{ij}$.

\subsection{\label{subsec:effect}Suitable effective counterterms for the gap equations}
The divergent part in Eq.~(\ref{eq:Delta_ij-isol})
will be removed by effective mass counter terms. 
We explain in Appendix~\ref{sec:2PPI} how 
the effective mass counterterms we are using here are related to 
standard mass and coupling constant counterterms 
in a standard counterterm Lagrangian. More precisely 
the 2PPI formalism establishes a one-to-one mapping 
between both counterterm approaches.

In order to have a $\delta \calm^2$ counterterm in the gap equations, 
we add to the effective action in Eq.~(\ref{eq:Gamma-2PPI}) 
a counterterm of the general form
\bea
\delta \calm^4&=&\delta\xi_{\phi\phi}\left(\calm^2_{\phi\phi}\right)^2
+\delta\xi_{\chi\chi}\left(\calm^2_{\chi\chi}\right)^2\nonumber
\\
&&+2\,\delta\xi_{\phi\chi}\left(\calm^2_{\phi\chi}\right)^2 
\label{eq:delta-M4}
\eea
With the introduced effective mass counterterms 
the renormalized gap equations take the form
\bea
\calm^2_{\mathrm{R},\,\phi\phi}(t)&=&m^2
+g^2\left(\chi^2(t)+\Delta_{\chi\chi}(t)\right)\nonumber
\\
&&-4g^2\, \delta\xi_{\chi\chi}\calm^2_{\mathrm{R},\chi\chi}(t)\\
\calm^2_{\mathrm{R},\,\chi\chi}(t)&=&-\lambda
v^2+g^2\left(\phi^2(t)+\Delta_{\phi\phi}(t)\right)\nonumber \\
&&+3\lambda\left(\chi^2(t)+\Delta_{\chi\chi}(t)\right)\\
&&-4g^2 \,
\delta\xi_{\phi\phi}\calm^2_{\mathrm{R},\,\phi\phi}(t)-12\lambda\,
\delta\xi_{\chi\chi}\calm^2_{\mathrm{R},\,\chi\chi}(t)\nonumber\\
\calm^2_{\mathrm{R},\,\phi\chi}(t)
&=&2g^2\left(\phi(t)\chi(t)+\Delta_{\phi\chi}(t)\right)\nonumber  \\
&&-8g^2\,\delta\xi_{\phi\chi}\calm^2_{\mathrm{R},\,\phi\chi}(t)\ .
\eea
By inserting $\Delta_{ij}(t)$ from Eq.~\eqn{eq:Delta_ij-isol} 
one can see that the gap equations become finite if we choose 
\bea
\delta\xi_{\phi\phi}=\delta\xi_{\chi\chi}=\delta\xi_{\phi\chi}
&=&-\frac{1}{64\pi^2}\left[L_\epsilon+1\right]\\
&=&-\frac{1}{64\pi^2}\left(\frac{2}{\epsilon}-\gamma+1+\ln
4\pi\right) \nonumber \ .
\eea
This choice of $\delta\xi$ corresponds to a $\overline{\mathrm{MS}}$ 
prescription. In particular, the renormalization scheme is mass independent.
Thus the system of renormalized gap equations 
is finally given by
\bea
\calm^2_{\mathrm{R},\,\phi\phi}(t)
&=&m^2+g^2\left(\chi^2(t)+\Delta_{\chi\chi,\mathrm{fin}}(t)\right)
\nonumber\\ 
&&+g^2C^{0}_{\chi\chi}\label{gap1ren}\\
&&+g^2
C^{1}_{\chi\chi}\left(\calm^2_{\mathrm{R},\,\phi\phi}(t)
-\calm^2_{\mathrm{R},\,\phi\phi}(0)\right)\nonumber\\ 
&&+g^2
C^{2}_{\chi\chi}\left(\calm^2_{\mathrm{R},\,\chi\chi}(t)
-\calm^2_{\mathrm{R},\,\chi\chi}(0)\right)\nonumber\\ 
&&+g^2
C^{3}_{\chi\chi}\left(\calm^2_{\mathrm{R},\,\phi\chi}(t)
-\calm^2_{\mathrm{R},\,\phi\chi}(0)\right)\nonumber
\ ,\\ 
\calm^2_{\mathrm{R},\,\chi\chi}(t)&=&-\lambda v^2+
g^2\left(\phi^2(t)+\Delta_{\phi\phi,\mathrm{fin}}(t)\right)
\nonumber\\ 
&&+g^2
C^{0}_{\phi\phi}\label{gap2ren}\\
&&+g^2C^{1}_{\phi\phi}\left(\calm^2_{\mathrm{R},\,\phi\phi}(t)
-\calm^2_{\mathrm{R},\,\phi\phi}(0)\right)\nonumber\\ 
&&+g^2
C^{2}_{\phi\phi}\left(\calm^2_{\mathrm{R},\,\chi\chi}(t)
-\calm^2_{\mathrm{R},\,\chi\chi}(0)\right)\nonumber\\ 
&&+g^2
C^{3}_{\phi\phi}\left(\calm^2_{\mathrm{R},\,\phi\chi}(t)
-\calm^2_{\mathrm{R},\,\phi\chi}(0)\right)\nonumber\\ 
&&+3\lambda
\left(\chi^2(t)+\Delta_{\chi\chi,\mathrm{fin}}(t)\right)
+3\lambda C^{0}_{\chi\chi}\nonumber \\
&&+3\lambda 
C^{1}_{\chi\chi}\left(\calm^2_{\mathrm{R},\,\phi\phi}(t)
-\calm^2_{\mathrm{R},\,\phi\phi}(0)\right)\nonumber\\ 
&&+3\lambda
C^{2}_{\chi\chi}\left(\calm^2_{\mathrm{R},\,\chi\chi}(t)
-\calm^2_{\mathrm{R},\,\chi\chi}(0)\right)\nonumber\\ 
&&+3\lambda
C^{3}_{\chi\chi}\left(\calm^2_{\mathrm{R},\,\phi\chi}(t)
-\calm^2_{\mathrm{R},\,\phi\chi}(0)\right)\nonumber \ , \\
\calm^2_{\mathrm{R},\,\phi\chi}(t)
&=&2g^2\left(\phi(t)\chi(t)+\Delta_{\phi\chi,\mathrm{fin}}(t)\right)
\nonumber\\ 
&&+2g^2 C^{0}_{\phi\chi}\label{gap3ren}\\
&&+2g^2 C^{1}_{\phi\chi}
\left(\calm^2_{\mathrm{R},\,\phi\phi}(t)
-\calm^2_{\mathrm{R},\,\phi\phi}(0)\right)\nonumber\\ 
&&+2g^2
C^{2}_{\phi\chi}\left(\calm^2_{\mathrm{R},\,\chi\chi}(t)
-\calm^2_{\mathrm{R},\,\chi\chi}(0)\right)\nonumber\\ 
&&+2g^2 C^{3}_{\phi\chi}\left(\calm^2_{\mathrm{R},\,\phi\chi}(t)
-\calm^2_{\mathrm{R},\,\phi\chi}(0)\right)\nonumber \ .   
\eea
This system of linear equations is similar to 
the one appearing in the $O(N)$-model in the Hartree approximation 
\cite{Baacke:2001zt}. It has to be
solved at each time. However, the coefficient matrix can be diagonalized 
with a time-independent rotation matrix, because it is 
time independent itself. Such a rotation matrix is analogous to 
the factor $\mathcal{C}=(1+\frac{\lambda}{16\pi^2}\ln\frac{m^2}{m_0^2})^{-1}$
in the renormalization of the $O(N)$-model in the 
large-$N$ approximation \cite{Baacke:2000fw}.


\subsection{\label{sec:ren-energy}Renormalized energy}
Within the Hartree approximation the contributions 
to the energy introduce logarithmic, 
quadratic and quartic divergences. 
The divergences have to be compensated by the already fixed 
counterterms of the last section.

If the effective masses are identified by the 
renormalized effective masses of the previous section, then
the zero-loop contribution to the energy [see Eq.~(\ref{eq:E-zero})]
is automatically renormalized.

In the following we use the rotated potential $\widetilde{\mathcal{V}}$ 
defined by
\bea
\widetilde{\mathcal{V}}_{\alpha\beta}(t)
&=&O_{k\alpha}\mathcal{V}_{kl}(t)O_{l\beta}\label{eq:def_vtilde}\\
&=&O_{1\alpha}O_{1\beta} \mathcal{V}_{\phi\phi}(t)
+O_{2\alpha}O_{2\beta}\mathcal{V}_{\chi\chi}(t)\nonumber \\
&&+(O_{1\alpha}O_{2\beta}+O_{2\alpha}O_{1\beta})\mathcal{V}_{\phi\chi}(t)\ 
\eea
According to the expansion of the mode functions in 
Appendix~\ref{sec:pert-expans} the divergent part of the 
one-loop contribution from the bubble graphs to the quantum energy 
is given by [see Eq.~(\ref{eq:E1_div})]
\bea
E^{(1),\,\mathrm{div}}(t)
&=&\frac{1}{2}\int\frac{d^{D-1}p}{(2\pi)^{D-1}}
\sum_\alpha\frac{1}{2\omega_\alpha}
\bigg[2\omega_\alpha^2+\widetilde{\mathcal{V}}_{\alpha\alpha}(t)
\tcsplit
-\sum_\beta
\frac{1}{2\omega_\beta(\omega_\alpha+\omega_\beta)}
\widetilde{\mathcal{V}}_{\alpha\beta}(t)\widetilde{\mathcal{V}}_{\alpha\beta}(t)\bigg] 
\ .\quad\qquad
\eea
The first term is quartic divergent. Its renormalization 
corresponds to a renormalization of the cosmological 
constant $\Lambda$; it is therefore somewhat arbitrary 
and can be omitted.

If the divergent parts are evaluated in dimensional regularization 
[using Eq.~(\ref{eq:fi-fish}),~(\ref{eq:fi-tadpole}) 
and (\ref{eq:quartic-div}) in Appendix~\ref{sec:feynman-iden}] the full 
one-loop contribution in Eq.~(\ref{eq:energy-E1}) denotes
\bea
E^{(1)}(t)&=&E^{(1)}_\mathrm{fin}(t)
+\sum_\alpha\frac{m_{0,\,\alpha}^4}{64\pi^2}
\left[L_\epsilon-\ln\frac{m_{0,\,\alpha}^2}{\mu^2}
+\frac{3}{2}\right]\nonumber
\\
&&-\sum_\alpha\frac{m_{0,\,\alpha}^2}{32\pi^2}
\widetilde{\mathcal{V}}_{\alpha\alpha}(t)
\left[L_\epsilon-\ln\frac{m_{0,\,\alpha}^2}{\mu^2}+1\right]\nonumber
\\
&&-\sum_{\alpha,\,\beta}\frac{1}{64\pi^2}
\widetilde{\mathcal{V}}_{\alpha\beta}(t)
\widetilde{\mathcal{V}}_{\alpha\beta}(t)
\tcsplit
\times
\bigg[L_\epsilon-\ln\frac{m_{0,\,\alpha}^2}{\mu^2}+1
\tcsplit
\qquad
+\frac{m_{0,\,\beta}^2}{m_{0,\,\alpha}^2-m_{0,\,\beta}^2}
\ln\frac{m_{0,\,\beta}^2}{m_{0,\,\alpha}^2}\bigg]  \ ,
\eea
where the finite part has been defined as
\bea
E^{(1)}_\mathrm{fin}(t)
&=&\frac{1}{2}\int\frac{d^{D-1}p}{(2\pi)^{D-1}}
\sum_\alpha\frac{1}{2\omega_\alpha}
\tcsplit
\, \times
\Bigg\{\mathrm{Re}
\left[\dot{f}_\phi^\alpha(t;p)\dot{f}_\phi^{\alpha\,*}(t;p)\right]
\tcsplit
\quad
+\mathrm{Re}\left[\dot{f}_\chi^\alpha(t;p)
\dot{f}_\chi^{\alpha\,*}(t;p)
\right]\nonumber \\
&&\quad
+\left(\vec{p}^2+\calm^2_{\phi\phi}(t)\right)
\mathrm{Re}\left[f_\phi^\alpha(t;p)
f_\phi^{\alpha\,*}(t;p)\right]\nonumber
\\
&&\quad+\left(\vec{p}^2+\calm^2_{\chi\chi}(t)\right)
\mathrm{Re}\left[f_\chi^\alpha(t;p)
f_\chi^{\alpha\,*}(t;p)\right]\nonumber
\\
&&\quad+2\calm^2_{\phi\chi}(t)\mathrm{Re}
\left[f_\phi^\alpha(t;p)f_\chi^{\alpha\,*}(t;p)\right]\nonumber
\\
&&\quad -2\omega_\alpha^2-\widetilde{\mathcal{V}}_{\alpha\alpha}(t)
\tcsplit
\quad
+\sum_\beta
\frac{1}{2\omega_\beta(\omega_\alpha+\omega_\beta)}
\widetilde{\mathcal{V}}_{\alpha\beta}(t)
\widetilde{\mathcal{V}}_{\alpha\beta}(t)\Bigg\}
\label{eq:E1_fin}
\ .
\eea
From the definition of $m_{0,\,\alpha}^2$ and 
$\widetilde{\mathcal{V}}_{\alpha\beta}(t)$ one can prove the identity 
\bea
&&\sum_\alpha
\bigg[m_{0,\,\alpha}^4
-2m_{0,\,\alpha}^2\widetilde{\mathcal{V}}_{\alpha\alpha}(t)
-\sum_\beta\widetilde{\mathcal{V}}_{\alpha\beta}(t)
\widetilde{\mathcal{V}}_{\alpha\beta}(t)
\bigg]
\tcsplit
=-\calm^2_{ij}(t)\calm^2_{ij}(t)\ ,
\eea
so that the divergent part becomes very simple. The full quantum energy 
$E^{(1)}$ is then given by
\bea
E^{(1)}(t)&=&E^{(1)}_\mathrm{fin}(t)-\frac{1}{64\pi^2} 
\left[L_\epsilon+1\right]
\calm^2_{ij}(t)\calm^2_{ij}(t)\nonumber \\
&&+\sum_\alpha
\frac{m_{0,\,\alpha}^4}{64\pi^2}
\left[-\ln\frac{m_{0,\,\alpha}^2}{\mu^2}+\frac{1}{2}\right]
\tcsplit
+\sum_\alpha
\frac{m_{0,\,\alpha}^2}{32\pi^2}\widetilde{\mathcal{V}}_{\alpha\alpha}(t)
\ln\frac{m_{0,\,\alpha}^2}{\mu^2}
\nonumber\\
&&
+\sum_{\alpha,\beta}\frac{1}{64\pi^2}
\widetilde{\mathcal{V}}_{\alpha\beta}(t)\widetilde{\mathcal{V}}_{\alpha\beta}(t)
\tcsplit
\times\bigg[\ln\frac{m_{0,\,\alpha}^2}{\mu^2}
-\frac{m_{0,\,\beta}^2}{m_{0,\,\alpha}^2
-m_{0,\,\beta}^2}\ln\frac{m_{0,\,\beta}^2}{m_{0,\,\alpha}^2}\bigg]
 \ .
\eea
The divergent part of $E^{(1)}$ is proportional to 
$\calm^2_{ij}\calm^2_{ij}=
\calm^4_{\phi\phi}+\calm^4_{\chi\chi}+2\calm^4_{\phi\chi}$,
i.e. the counterterm in Eq.~(\ref{eq:delta-M4}), 
as it has been expected.

Within the given approximation we can write down the renormalized 
total energy as
\bea
E_\mathrm{tot}&=&E^{(0)}(t)+E^{(1)}(t)-\delta \calm^4 \\ 
&=&E^{(0)}(t)\nonumber +E_\mathrm{fin}^{(1)}(t)+\sum_\alpha
\frac{m_{0,\,\alpha}^4}{64\pi^2}
\left[-\ln\frac{m_{0,\,\alpha}^2}{\mu^2}+\frac{1}{2}\right]\nonumber
\\
&&+\sum_\alpha
\frac{m_{0,\,\alpha}^2}{32\pi^2}\widetilde{\mathcal{V}}_{\alpha\alpha}(t)
\ln\frac{m_{0,\,\alpha}^2}{\mu^2}
\nonumber\\
&&
+\sum_{\alpha,\beta}\frac{1}{64\pi^2}
\widetilde{\mathcal{V}}_{\alpha\beta}(t)\widetilde{\mathcal{V}}_{\alpha\beta}(t)
\tcsplit
\times
\bigg[\ln\frac{m_{0,\,\alpha}^2}{\mu^2}
-\frac{m_{0,\,\beta}^2}{m_{0,\,\alpha}^2
-m_{0,\,\beta}^2}\ln\frac{m_{0,\,\beta}^2}{m_{0,\,\alpha}^2}\bigg] \ .
\eea
Because we have added the counterterm $\delta\calm^4$ to the effective 
action $\Gamma$, i.e., $\Gamma_\mathrm{R}=\Gamma+\delta \calm^4$, 
it has a negative sign in the renormalized energy.


\subsection{Renormalized equations of motion}

In summary one has to solve in the Hartree approximation the following 
renormalized 
equations of motion numerically. 
The classical equations of motion are given by
\bea
0&=&\ddot{\phi}(t)+\calm^2_{\mathrm{R},\,\phi\phi}(t)\phi(t)
+\calm^2_{\mathrm{R},\,\phi\chi}(t)\chi(t)
\tcsplit
-2g^2\chi^2(t)\phi(t)
\label{eq:phi-eq-ren}\ , \\
0&=&\ddot{\chi}(t)+\calm^2_{\mathrm{R},\,\chi\chi}(t)\chi(t)
+\calm^2_{\mathrm{R},\,\phi\chi}(t)\phi(t)
\tcsplit
-2\lambda \chi^3(t)-2g^2\phi^2(t)\chi(t) \label{eq:chi-eq-ren}\ , 
\eea
while the equations for the mode functions denote explicitly
\bea
0&=&\ddot{f}_\phi^\alpha (t;p)+\vec{p}^2
f_\phi^\alpha(t;p)
\tcsplit
+\,\calm^2_{\mathrm{R},\,\phi\phi}(t) f_\phi^\alpha(t;p)
+\calm^2_{\mathrm{R},\,\phi\chi}(t) f_\chi^\alpha(t;p)
\label{eq:fphi-eq-ren} \\ 
0&=&\ddot{f}_\chi^\alpha (t;p)+\vec{p}^2
f_\chi^\alpha(t;p)
\tcsplit
+\,\calm^2_{\mathrm{R},\,\chi\chi}(t) f_\chi^\alpha(t;p)
+\calm^2_{\mathrm{R},\,\phi\chi}(t) f_\phi^\alpha(t;p)
\label{eq:fchi-eq-ren}  \ .
\eea
In addition one has to solve, at each time, the $3\times 3$ system
of renormalized gap equations \eqn{gap1ren}--\eqn{gap3ren}
for the masses $\calm^2_{ij}$.


\section{Numerical Implementation}
\label{sec:numer-impl}
The masses $m_{0,\,\alpha}^2$ ($\alpha=1,2$) and the mixing 
angle $\vartheta$ have to be determined self-consistently 
at the initial time $t=0$. 
The renormalization scale $\mu$ for the finite parts is 
fixed to $\mu^2=\lambda v^2$. 

The equations of motion~\eqn{eq:phi-eq-ren}--\eqn{eq:fchi-eq-ren} 
are solved using a standard fourth order 
Runge-Kutta algorithm.
We use a time discretization $\Delta t=0.0003$. 
The momentum integrations are carried out on a momentum grid 
with a non-equidistant momentum discretization. 
We use a pragmatic momentum cutoff $p_\mathrm{max}=12$ and $n_p=300$ momenta
for the convergent fluctuation integrals in 
Eq.~(\ref{eq:Delta_fin})~and~(\ref{eq:E1_fin}).
The accuracy of the numerical computations is monitored by verifying 
the constancy of the Wronskians and of the total energy.

\section{Results}
\label{sec:results}

\subsection{General outline}
The basic conception in the Hybrid model is an efficient energy
transfer from the inflaton to the Higgs degree of freedom mediated
by a phase transition and the associated spinodal regime.
The Lagrangian is constructed in such a way that this type
of behavior can be expected. It is then a question
how these expectations are realized; as mentioned in the Introduction
this has been investigated in various approximations
on the classical or quantum level. Here we present numerical
simulations in 
the Hartree approximation, which encompasses spatially homogeneous
classical fields, and quantum fluctuations with a specific
back reaction among themselves. 
Specifically we are interested 
\begin{enumerate}
\item to see on which time scale and in which form the energy transfer 
between the inflaton and Higgs
fields takes place
\item
to conclude on the structure of the effective Higgs potential 
after this energy transfer.
Though the system does not go right away into a thermal equilibrium
phase the behavior at intermediate and late times can be 
thought as reflecting the shape of an effective Higgs potential, with a 
symmetric or broken symmetry structure. 
\item
in the spectra of the different quantum modes reflecting the
mechanism of particle production
\item
in finding out to which extent the transition to a 
classical description may be justified in a certain momentum
range.
\end{enumerate}

\begin{figure}[htbp]
\begin{center}
\includegraphics[width=8.15cm]{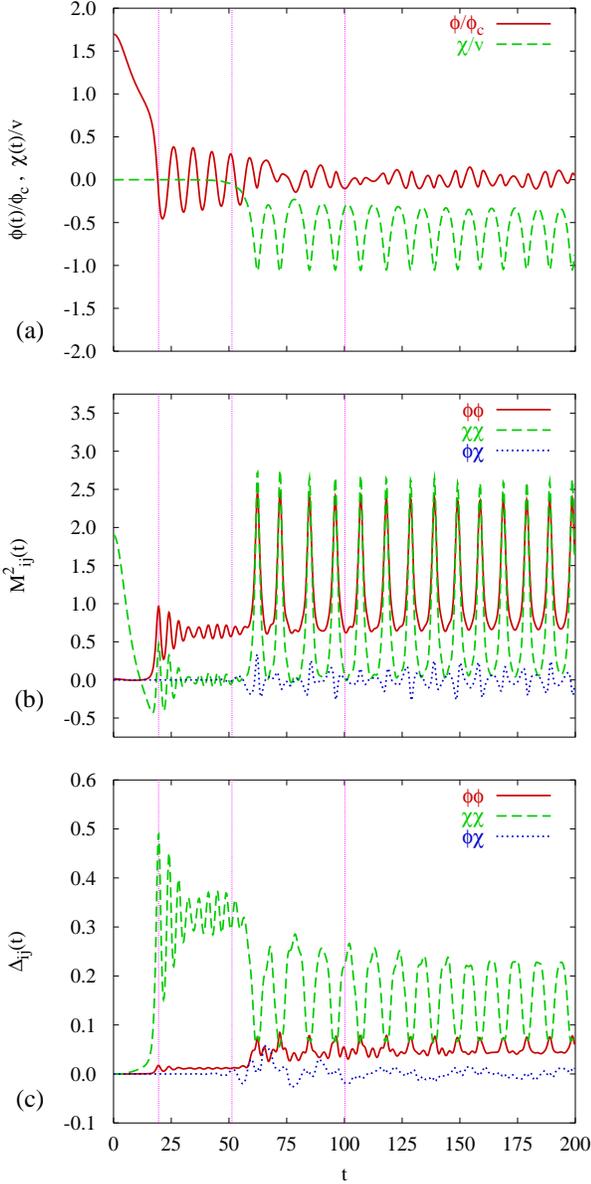}
\end{center}
  \caption{(Color online) 
Time evolution for the simulation with $g^2=2\lambda$. 
Initial values: $\phi(0)=1.2$ and 
$\chi(0)=1.0\times 10^{-7}$; other parameters: $m^2=0$, $\lambda=1$, $v^2=1$; 
we plot as a function of time
 (a) the classical fields 
$\phi(t)/\phi_\mathrm{c}$ (red solid line) 
and $\chi(t)/v$ (green dashed line), 
(b) the effective masses $\calm^2_{ij}(t)$ with $ij={\phi\phi}$ 
(red solid line), 
$ij={\chi\chi}$ (green dashed line) and $ij={\phi\chi}$ 
(blue dotted line)
(c) the fluctuation integrals $\Delta_{ij}(t)$ with 
$ij={\phi\phi}$ (red solid line), 
$ij={\chi\chi}$ (green dashed line) and $ij={\phi\chi}$ 
(blue dotted line); the vertical dotted lines indicate 
the times where $t$ is equal to $19.5$, $51.3$ and $100.2$}
\label{fig:Set-5-2}
\end{figure}

\begin{figure}[htbp]
\begin{center}
\includegraphics[width=8.15cm]{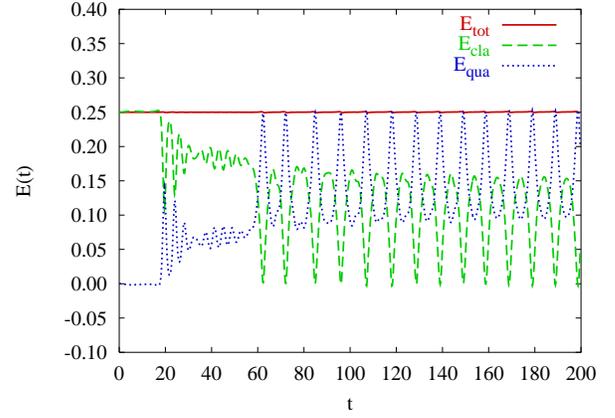}
\end{center}
  \caption{(Color online)
 Energy contributions for the simulation with parameters in 
Fig.~\ref{fig:Set-5-2}; the red solid line denotes the total energy, 
the green dashed line the classical energy and the blue dotted 
line the quantum energy}
  \label{fig:energy-2}
\end{figure}

\begin{figure}[htbp]
\begin{center}
\includegraphics[width=8.15cm]{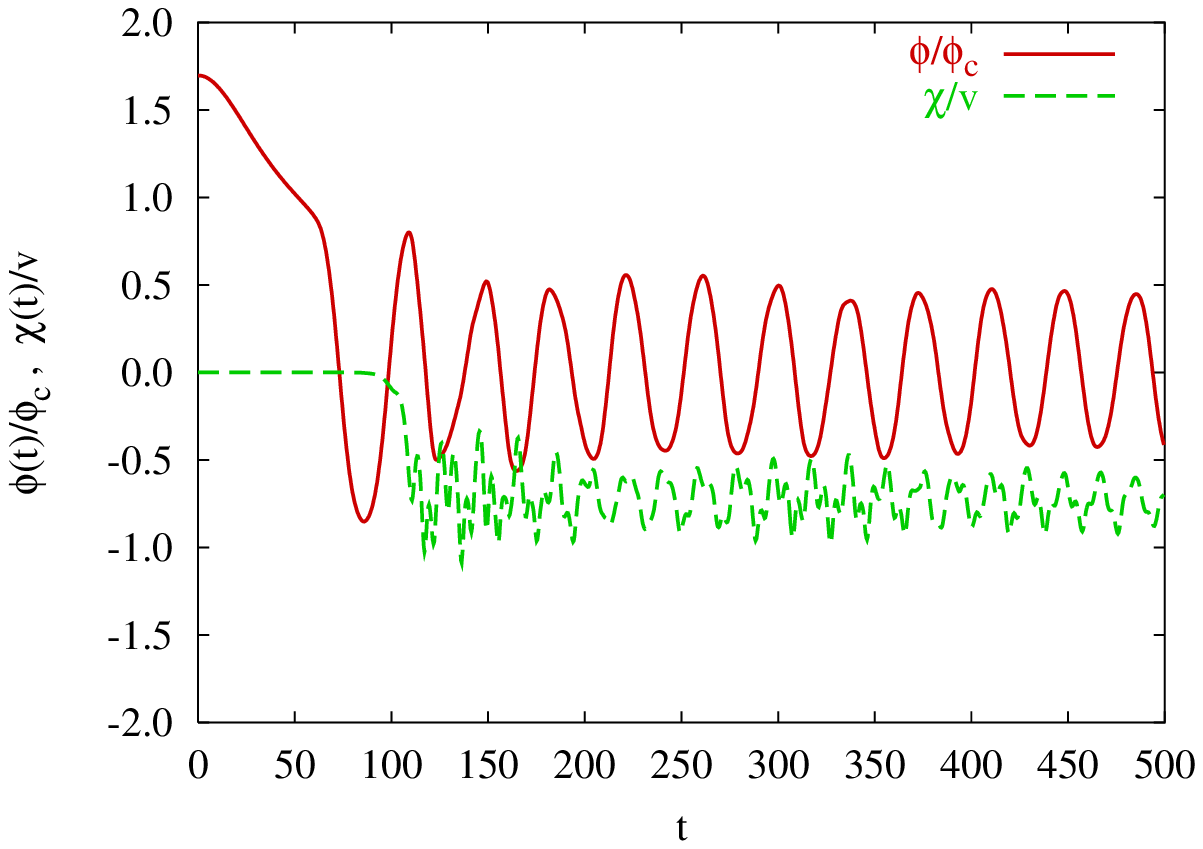}
\end{center}
  \caption{(Color online)
 The same as Fig.~\ref{fig:Set-5-2}a but for the 
coupling case $g^2=0.1\lambda$ and the initial value $\phi(0)=5.366$}
  \label{fig:Set-5-3}
\end{figure}

\begin{figure}[htbp]
\begin{center}
\includegraphics[width=8.15cm]{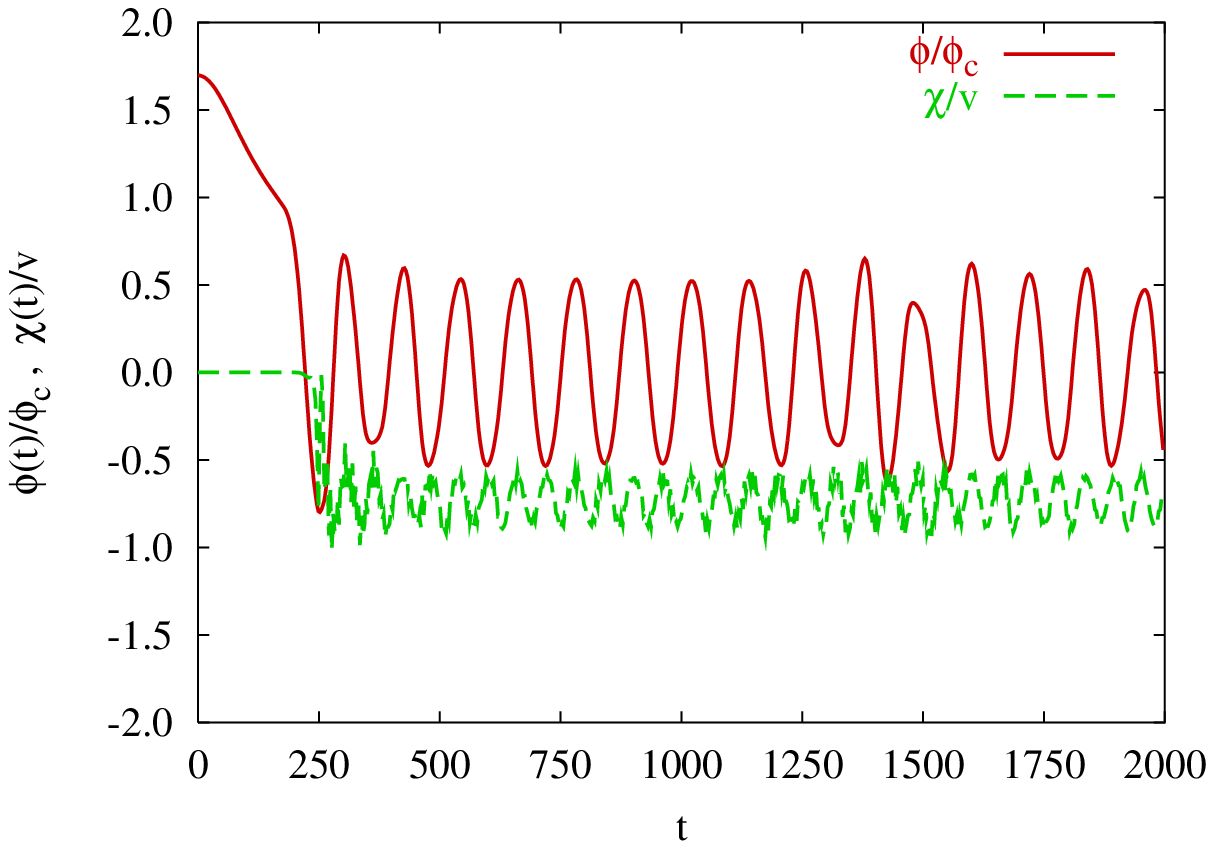}
\end{center}
  \caption{(Color online)
 The same as Fig.~\ref{fig:Set-5-2}a but for the 
coupling case $g^2=0.01\lambda$ and the initial value $\phi(0)=16.97$}
\label{fig:Set-5-1}
\end{figure}

\begin{figure}[htbp]
\begin{center}
\includegraphics[width=8.15cm]{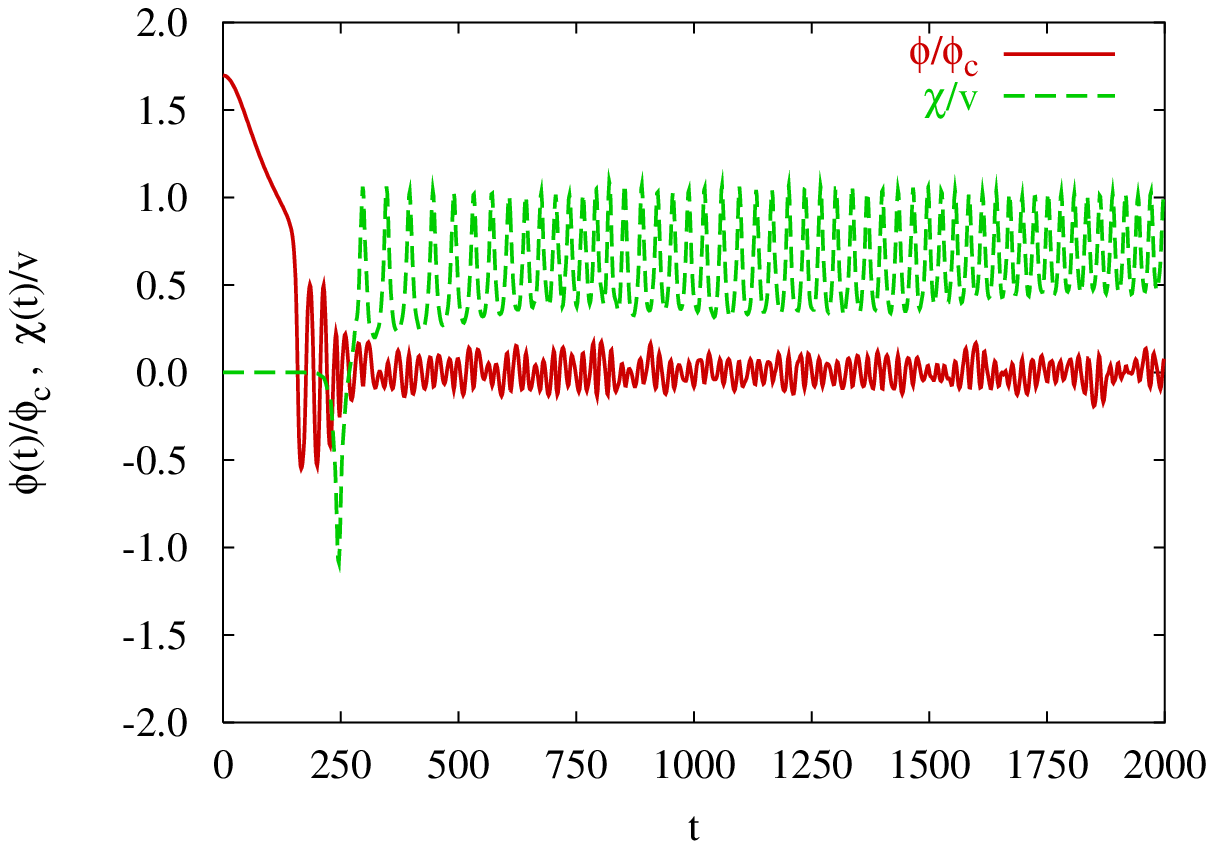}
\end{center}
  \caption{(Color online)
 The same as Fig.~\ref{fig:Set-5-2}a but for $\lambda=0.1$}
\label{fig:Set-5-5}
\end{figure}

The answer to these questions obviously depends on the parameters
chosen for the simulations.
In order to study the influence of the coupling strength $g^2$ and the 
self-coupling $\lambda$ on both the classical and quantum components of the 
Higgs and the inflaton fields we have performed simulations with 
$m^2=0$, $v^2=1$, $\lambda=1$ fixed, while $g^2$ is equal to 
$2\lambda$, $0.1\lambda$ and $0.01\lambda$ (Fig.~\ref{fig:Set-5-2}, 
\ref{fig:Set-5-3} and \ref{fig:Set-5-1}) and for $g^2=2\lambda$ with a 
smaller coupling $\lambda=0.1$ (Fig.~\ref{fig:Set-5-5}). 
We have chosen $\chi(0)=10^{-7}$, i.e., a very small value for the initial 
amplitude of the classical Higgs field, in order to trigger
the ``spontaneous'' symmetry breaking. 
The initial amplitude for the inflaton field has been fixed for all 
cases to $\phi(0)=1.697\phi_\mathrm{c}$. The energy contributions for 
the simulation in Fig.~\ref{fig:Set-5-2} are displayed in 
Fig.~\ref{fig:energy-2}.

\subsection{Time regimes and exponential growth}
\label{sec:time-regimes}
We have identified three time regimes in the simulations 
that we want to investigate further in the following.
The regimes are described as follows:
\begin{list}{}{}
\item[(I)] Initial period,  end of slow roll.
Here $\calm^2_{\chi\chi}(t)>0$. A phase of slow rolling 
of the inflaton field after the main period of inflation. 
The quantum fluctuations still are almost negligible. 
\item[(II)] Early times, spinodal regime, $\calm^2_{\chi\chi}(t)<0$
or oscillating several times around zero. Spinodal amplification of 
Higgs quantum fluctuations and and exponential growth of $\chi(t)$.
\item[(III)] Intermediate and late times, $\calm^2_{\chi\chi}(t)>0$ 
and oscillations of the classical fields. Excitation of inflaton and 
mixed quantum fluctuations, parametric resonance bands in all momentum
spectra.
\end{list}
The first period (I) is easy to identify in
 Figs.~\ref{fig:Set-5-2}--\ref{fig:Set-5-3}: only the inflaton
decreases with time in  smooth way while the Higgs mean field is still
practically zero. 

\begin{figure}[htbp]
\centering
\includegraphics[width=8.15cm]{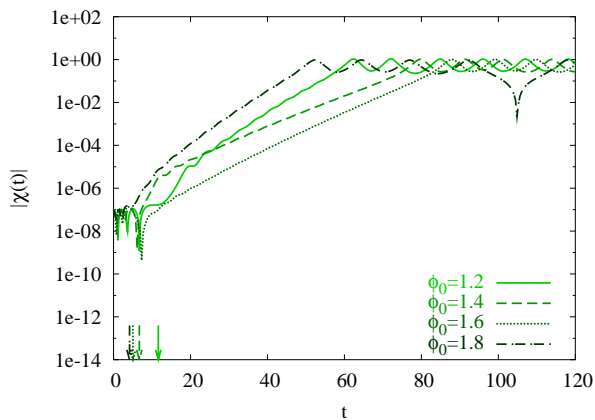}
\caption{Time evolution of the absolute value $|\chi(t)|$ 
for simulations with the parameters $g^2=2\lambda$, 
$m^2=0$, $v^2=1$ and $\lambda=1$ and four different initial values 
$\phi(0)$ equal to $1.2$ (solid line), 
$1.4$ (dashed line), $1.6$ (dotted line) and $1.8$ (dashed-dotted line); 
the corresponding arrows pointing at the $t$-axis indicate the time 
when $\phi(t)$ drops below $\phi_\mathrm{c}$}
\label{fig:chiabs}
\end{figure}

In the  early time period (II) the inflaton field passes
through zero once or several times, depending on the
coupling $g^2$, see Figs.~\ref{fig:Set-5-2}--\ref{fig:Set-5-1}. 
The period is identified by an
increase of $|\chi(t)|$ and ends once $\chi(t)$ begins to
oscillate in a regular way. 
A closer analysis shows that the amplitude of the classical Higgs field 
growths exponentially. In Fig.~\ref{fig:chiabs} we display 
on a logarithmic scale the absolute value $|\chi(t)|$ 
for simulations with $g^2=2\lambda$, 
$\lambda=1$, $m^2=0$, $v^2=1$ and $\chi(0)=10^{-7}$ fixed, 
while the initial amplitude of the inflaton field, $\phi(0)$, 
is varied from $1.2$ to $1.8$. 
The exponential growth sets in 
when $\phi(t)$ becomes smaller than the critical value $\phi_\mathrm{c}$ 
(see Eq.~(\ref{eq:phi_c})) and stops when $\chi(t)$ 
reaches the turning point which is at $|\chi(t)|\approx 1$. 
There does not seem to be a systematic trend for 
the dependence of the period of growth  on $\phi(0)$.

The regime (II) can be very short. For the simulation with a small 
coupling $g^2=0.01\lambda$  the transition to the broken symmetry phase 
can take place within a single oscillation of the inflaton field 
(see Fig.~\ref{fig:Set-5-1}a). 

The intermediate and late time period (III) is characterized by
oscillations of both the inflaton and the Higgs field, with essentially
constant period and amplitude (see, e.g., Fig.~\ref{fig:Set-5-5}). 
The Higgs mean field may oscillate around
a nonzero value, related to a broken symmetry minimum of an 
effective potential (see below) or around $\chi=0$, to be identified
with symmetry restoration. This period is further analyzed in the 
next subsection.

\subsection{Late time averages -- phase transition}
\label{sec:late-time-averages}
The amplitudes of the classical fields $\phi$ and $\chi$ decrease
very slowly, if at all, at late times, i.e.,  once they have started to 
oscillate in a kind of effective potential. Though we make no attempt
to reconstruct such a potential in detail, the oscillations allow
to conclude on the minimum and the range of such an effective potential
for both the Higgs and inflaton fields. In this sense we can 
speak of a symmetric or broken symmetry phase for the Higgs
field, if the minimum of its effective potential is at
$\chi=0$ and $\chi \neq 0$, respectively, and we associate this 
minimum with the time average of the Higgs field at late
times. The shape of the effective potential depends here on the
energy density (in place of the temperature) and therefore
on the initial value of the inflaton field.
The question of spontaneous symmetry breaking and of the point of
the phase transition reduces therefore to finding the
the dependence of $\chi(t\to \infty)$ on the initial value 
$\phi(0)$. 

\begin{figure}[htbp]
  \centering
  \includegraphics[width=8.15cm]{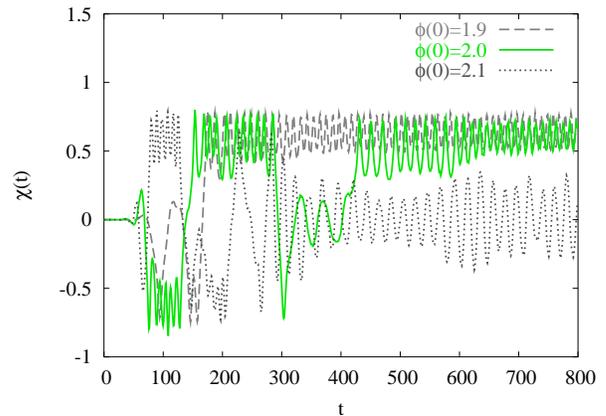}
  \caption{(Color online) Time evolution of $\chi(t)$ with parameters as 
in Fig.~\ref{fig:chiabs} 
but for the initial values $\phi(0)$ equal to $1.9$ (dashed line), 
$2.0$ (green solid line) and $2.1$ (dotted line).}
\label{fig:chi_inf-time}
\end{figure}

In order to study this issue we have performed a series of simulations 
where we have varied the initial amplitude $\phi(0)$ while keeping
all the other parameters fixed. 
In Fig.~\ref{fig:chi_inf-time} the time evolution of $\chi(t)$ for simulations 
with $\phi(0)=1.9$, $2.0$ and $2.1$ is displayed. The other parameters are
$g^2=2\lambda$, $\lambda=1$, $v=1$, $m=0$ and $\chi(0)=10^{-7}$.

A first inspection suggests that there is a phase transition 
between $\phi(0)=2.0$ and $\phi(0)=2.1$, 
as for the latter simulation the field $\chi$ oscillates around zero.
From the simulation with $\phi(0)=2.0$ 
in Fig.~\ref{fig:chi_inf-time} it becomes apparent that the field $\chi$ 
can jump several times from one ``minimum'' to the other if $\phi(0)$ is 
close to the critical point of the phase transition. This is typical for a 
first-order phase transition and has been observed in the scalar $O(N)$ 
model in the Hartree approximation as well \cite{Baacke:2001zt}.

\begin{figure}[htbp]
  \centering
  \includegraphics[width=8.15cm]{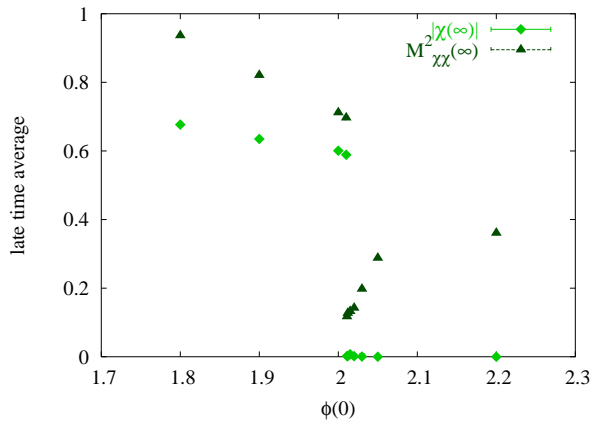}
  \caption{The late-time absolute value $|\chi(\infty)|$ (diamonds) 
and the late
  time effective mass $\calm^2_{\chi\chi}(\infty)$ (triangles) averaged at 
times $t\approx 1000$ as a function of the initial amplitude $\phi(0)$; 
other parameters as in Fig.~\ref{fig:Set-5-2}.}
  \label{fig:chi_inf-av}
\end{figure}

In Fig.~\ref{fig:chi_inf-av} we display the time averages 
$|\overline{\chi}(\infty)|$ and $\overline{\calm^2}_{\chi\chi}(\infty)$ 
as a function of the initial amplitude $\phi(0)$. 
The other parameters are fixed to $g^2=2\lambda$, $\lambda=1$, $v=1$, 
$m=0$ and $\chi(0)=10^{-7}$.
A first-order phase transition is signaled by a non-continuous drop of the 
minimum value $|\overline{\chi}(\infty)|$ and the effective mass 
$\calm^2_{\chi\chi}$ from finite (positive) values to zero.

\subsection{Momentum spectra}
\label{sec:momentum-spectra}

Using the amplitudes $f_j^\alpha(t,p)$ one may define 
various ``power spectra''. One of those is the integrand
of the fluctuation integrals $\Delta_{ij}(t)$, i.e., the tadpole
contributions. We have already introduced the kernel
\be
G_{ij}(t,t,\bfp)=\langle \tilde\Phi^*_i(t,\bfp)
\tilde \Phi(t,\bfp)\rangle/V
\ee
in terms of which we define the power spectrum of the fluctuation
amplitudes
\bea
P_{ij}(t,p)= G_{ij}(t,t,\bfp)\frac{p^2}{2\pi^2}
\label{eq:fluctuation-spectra}
\eea 
by including the momentum phase space factor.

\begin{figure*}[htbp]
\begin{center}
\includegraphics[width=17cm]{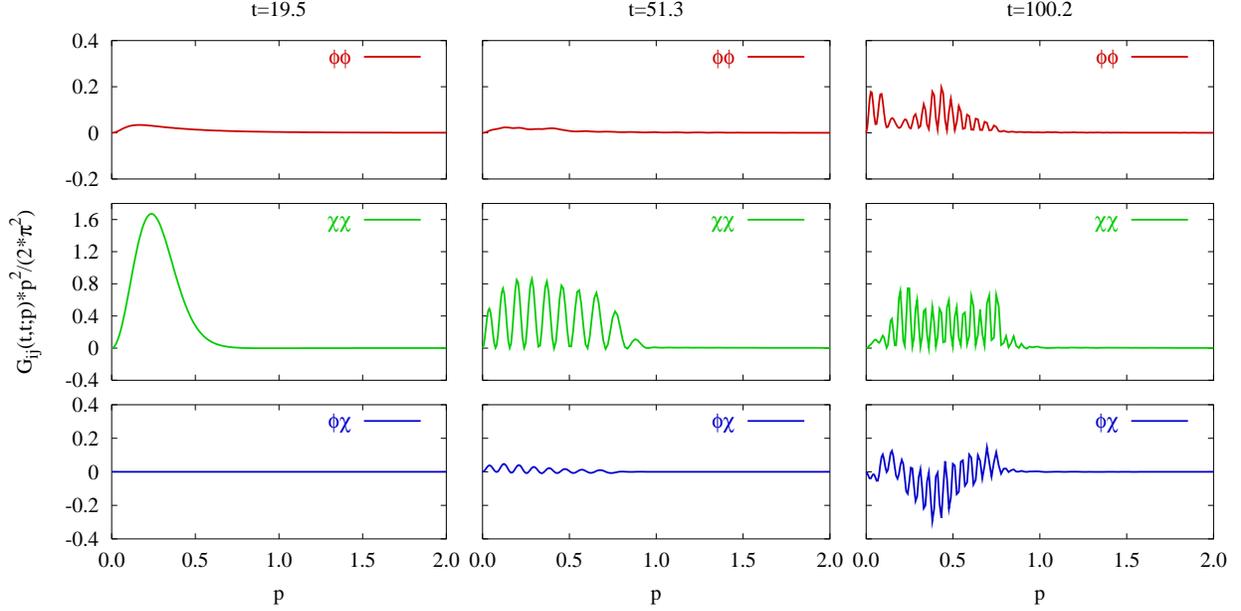}
\end{center}
\caption{Momentum spectra $G_{ij}(t,t;\vec{p})p^2/(2\pi^2)$ 
for the simulation in Fig.~\ref{fig:Set-5-2} at the times $t=19.5$ (left), 
$t=51.3$ (middle) and $t=100.2$ (right)}
\label{fig:Set-5-2-spectra}
\end{figure*}

In Fig.~\ref{fig:Set-5-2-spectra} 
we display this spectrum for the simulation in Fig.~\ref{fig:Set-5-2} 
at the times $t$ equal to $19.5$, $51.3$ and $100.2$. 
These time steps are indicated in Fig.~\ref{fig:Set-5-2} by 
vertical dashed lines. Actually we have subtracted the free field
part and the first order perturbative part of this kernel, in analogy
to the right hand side of Eq. \eqn{eq:Delta_fin}. The free field part
rises linearly  with momentum; for $p=2$, the maximal value
used in our plots, it is has typical values of $0.05$ and would be
visible. It makes the tadpole integrals divergent; as discussed
above, in our computations this divergence is absorbed by dimensional 
regularization and renormalization. 

At early and intermediate times the Higgs fluctuations dominate. The 
inflaton and mixed fluctuation spectra only appear in the late-time
regime, are however subdominant even there.
At early times, $t < 20$ the Higgs spectrum is generated by
negative squared masses as in tachyonic preheating or quench
scenarios. In Ref. \cite{Garcia-Bellido:2002aj} it was found that
the peak in the momentum spectrum $p P_{\chi\chi}(t,p)$ can be fitted 
by a Gaussian; 
we similarly find, at $t=19.5$, a spectrum
\be 
p P_{\chi\chi}(t,p)\simeq A\exp
\left[-B (|\vec{p}|-C)^2\right]
\ee
with
\begin{eqnarray}
A&=&0.4482\pm 0.0038\\
B&=&36.1115\pm 0.7131\\
C&=&0.307995\pm 0.001168 
\end{eqnarray}

At intermediate times the smooth peak broadens and decays into spikes, 
typical of parametric resonance. Parametric resonance also dominates
the shape of the spectra at late times. As the period and amplitude
of oscillation change very slowly, the width of the spectrum
remains constant.


\subsection{Correlations}
\label{sec:correlations}
\begin{figure}[htbp]
  \centering
  \includegraphics[width=8.15cm]{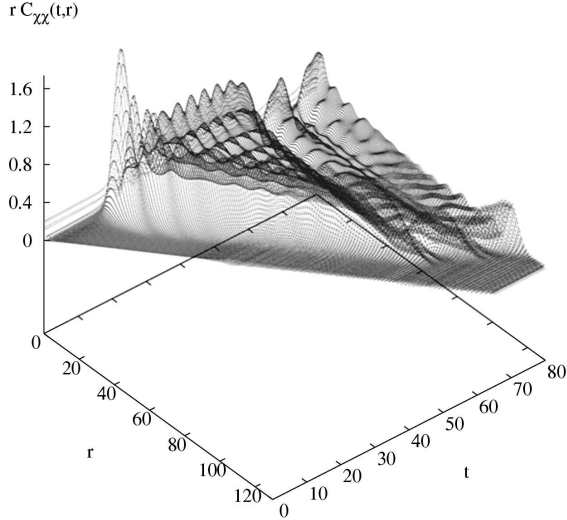}
  \caption{Correlation function $rC_{\chi\chi}(r,t)$ 
for the simulation in Fig.~\ref{fig:Set-5-2}}
  \label{fig:C_chichi}
\end{figure}

We define the correlation function between the different fluctuations
as
\bea
C_{ij}(r,t)&=&\int\frac{d^3p}{(2\pi)^3}e^{i\bfp\cdot \vec{x}}
\sum_\alpha \frac{1}{2\omega_\alpha} 
\Re\left(f_i^\alpha(t,p)f_j^{*\,\alpha}(t,p)\right)\nonumber\\
&=&\frac{1}{2\pi^2\, r} \int_0^\infty dp p \sin(pr)
\tcsplit
\quad\times
 \sum_\alpha \frac{1}{2\omega_\alpha} 
\Re\left(f_i^\alpha(t,p)f_j^{*\,\alpha}(t,p)\right)
\pkt\eea
We here consider the correlations of the Higgs fluctuations ($i=j=2$), 
which are displayed in Fig.~\ref{fig:C_chichi}. 
We observe the correlations
to be positive and propagating with $\Delta r = 2 \Delta t$, as
also found in the large-$N$ approximation \cite{Boyanovsky:1998yp}.
The propagation with twice the speed of light can be related
to the fact that the quantum fluctuations are correlated by the mean fields
whose influence propagates in opposite space directions.
This is corroborated by a strong decrease of such correlations
when the mean field amplitude goes to zero \cite{Baacke:2001zt}.
 
\subsection{Decoherence time}
\label{Decoherencetime}
\begin{figure}[htbp]
  \centering
  \includegraphics[width=8.15cm]{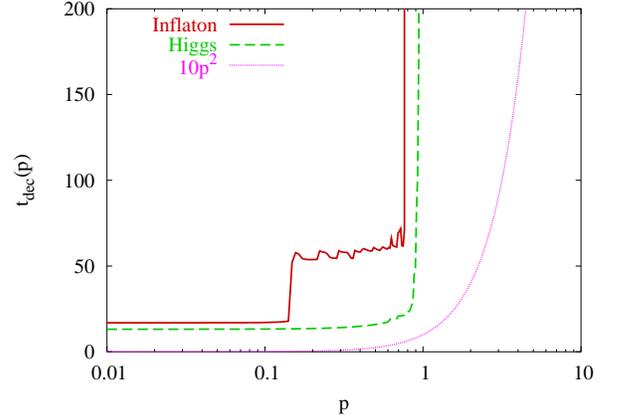}
  \caption{(Color online) 
Decoherence time $t_\mathrm{dec}(p)$ for which a given mode 
$p$ becomes ``classical'' ($|F_{ij}(t_\mathrm{dec},\bfp)|=1$) 
for the simulation in Fig.~\ref{fig:Set-5-2}; 
the red solid line represents the inflaton ($i=j=1$) and the 
green dashed line the 
Higgs modes ($i=j=2$); the dotted line corresponds to 
$t_\mathrm{dec}\propto p^2$.}
  \label{fig:tdec-p}
\end{figure}

One of the important questions is the justification of using
classical instead of quantum dynamics. In the context of nonequilibrium
quantum field theory this has been  discussed in Refs.
\cite{Guth:1985ya,Polarski:1996jg,Khlebnikov:1996mc} and applied to the
hybrid model in Refs. \cite{Asaka:2001ez,Garcia-Bellido:2002aj}. We use here,
adapted to our normalization, the definitions of 
Ref. \cite{Garcia-Bellido:2002aj}.
The ``classicality'' is measured by the imaginary part $F(t,\bfp)$ of a
correlation function:
\be
F_{ij}(t,\bfp)=\im \left[\sum_\alpha \frac{f_i^{\alpha *}(t,p)
\dot f_j^\alpha(t,p)}{2\omega_\alpha}\right]
\ee
the real part of the bracket being associated with the commutator.
The criterium for a classical description is given by 
\be
|F_{ii}| \gg 1
\pkt\ee
We display in Fig. \ref{fig:tdec-p} the time for the onset of classicality 
(``decoherence time'') as a 
function of momentum.

As far as the Higgs fluctuations ($i=j=2$) are concerned 
the figure can be compared to those of Refs. 
\cite{Asaka:2001ez,Garcia-Bellido:2002aj}. 
These authors consider the creation of quantum fluctuations
via the spinodal instability. They simplify the time evolution
by assuming for the mass of the Higgs fluctuations a behavior 
$\calm^2_{\chi\chi}\propto (t_0-t)$ where $t_0$ marks the onset
of the spinodal regime, without considering any back reactions
(see also 
Refs~\cite{Guth:1985ya,Bowick:1998kd,Copeland:2002ku,Lombardo:2002wt}).
In this case the mode functions become Airy functions and the results
are in analytic form. The boundary $|F(t,\bfp)|=1$ between the classical
and quantum regimes then behaves roughly \cite{Garcia-Bellido:2002aj} 
as $t\propto \bfp^2$. As displayed in Fig. \ref{fig:tdec-p}
 the shape of this boundary is quite different in our simulations, 
the classical regimes remains
limited within a fixed momentum band at all times. 
When one includes back reaction the behavior of the mass term
is linear in time only in a very limited time interval;
furthermore, due to the inflaton oscillations, the process repeats
several times. The limited momentum band for which
the modes can be considered as classical, can be seen as a 
consequence of parametric resonance. Due to the lack 
of strong dissipation the oscillations
of the classical fields persist at late times, and therefore 
also the resonance band. Whether and for which time period this is
physical or unphysical cannot be determined within the 
approximation used here, even though  it is certainly more
elaborate as previous approaches.

For the inflaton fluctuations,
not considered in Ref.\cite{Asaka:2001ez,Garcia-Bellido:2002aj}, 
the structure of this curve shows that 
fluctuations at very small
momenta become classical as early as those of the Higgs field, 
while those for larger momenta develop at later times. This 
is due, presumably, to a stronger role of parametric resonance
for the evolution of inflaton fluctuations. Again the band of
momenta for the classical regime remains sharply cutoff
even at late times.


\section{Conclusions}
\label{sec:conclusions}

In this final section we would like to comment
on some of our results and draw some conclusions.

On the formal level we have addressed the problem of renormalization
for a fully coupled system of two quantum fields in the Hartree
approximation, using methods similar to those in 
Refs.~\cite{Cormier:2001iw},\cite{Baacke:1997kj} 
and \cite{Baacke:2001zt}, but going beyond these applications. 
We consider this as an essential achievement of our work, which will
important when extending the model by including Goldstone and gauge fields
and the coupling to gravity where proper renormalization is
indispensable \cite{Baacke:1999gc}. Nevertheless we do not want
to infer that working with judiciously chosen cutoffs or less
elaborate schemes one may not reach sensible physical insights. 

When treated in the Hartree approximation this model lacks an 
efficient mechanism for dissipation.
This is considered to be a general drawback of the Hartree
approximation. The situation would improve, however, even in this
approximation, if Goldstone modes were included (see, e.g.
Fig.~1 in Ref.~\cite{Baacke:2001zt}). Dissipation via particle 
production is also found in the large-$N$ limit \cite{Boyanovsky:1995me}.
In more realistic models dissipation may proceed in addition
via fermion and gauge fields. The higher order effects
(sunset graphs or NLO-1/N) become effective only
when quantum fluctuations have grown to sufficient size.
This depends of course on the parameters of the model.
In the earlier stages the Hartree approximation should be
able to provide reliable information on the evolution
of decay and resonance processes. The time scales for thermalization
are expected to be much larger.

The model displays the expected transition to a broken symmetry
phase if the initial inflaton amplitude is not too large.
A possible intermediate restoration of symmetry by quantum
fluctuations would be followed by a later transition to the
broken symmetry state after further cosmological expansion.
Possible consequences like the unwanted formation of topological
defects have been discussed in the literature
\cite{Khlebnikov:1998sz:Parry:1998de}.

We have calculated the boundaries between regions where
a quantum description is needed and those where one may have recurse
to classical evolution equations. We find marked differences with
respect to previous work \cite{Asaka:2001ez,Garcia-Bellido:2002aj},
 where the production of quanta is
described in a simplified way, using a squared mass of the Higgs
field passing linearly through zero. We find numerically that the back 
reaction limits the classical regime to a low-momentum region
fixed for all times. 
Though large excitations of quantum fluctuations seem to justify the transition
to a classical description one has to keep in mind, that finally
one wants to end up with an ensemble described by quantum statistics
which is used in the standard thermal history of the early universe.
The classical ensembles suffer from the Rayleigh-Jeans divergence
incompatible with the finite amount of initial energy density. This 
automatically forces the fluctuations back to the quantum regime. 

The model discussed here corresponds to the original 
proposal by Linde, using a double well potential. This has been
used, as a simplification or a generic feature in some previous
studies \cite{Asaka:2001ez}. For electroweak
or GUT scale preheating the Higgs sector is based on a
symmetry group like $SU(n)$ or $SO(n)$ and will 
in general have more ($n_H$) degrees
of freedom. Near the spinodal point their masses are degenerate,
and this fact has been used in some studies
\cite{Garcia-Bellido:2002aj} in the way of just 
using $n_H$ identical copies of one and 
the same degree of freedom.
However, once the mean value of the Higgs field departs from zero,
there will be a nontrivial mass matrix for the quantum fluctuations
with several massless degrees of freedom, the would-be Goldstone bosons.
It can be expected that this will modify the quantum back reaction 
in an essential way. While this goes beyond the scope of the present
investigation, there are some studies using classical dynamics with 
more realistic Higgs sectors 
\cite{Felder:2001kt,Felder:2000hj,Borsanyi:2002tm:Borsanyi:2003ib}. 
Our formalism allows for a generalization towards more realistic
Higgs sectors, albeit with the limitation of homogeneous background
fields.

To become even more realistic, in particular in view of low scale
inflation, it would be very desirable to include gauge fields as well.
Unfortunately any resummation or backreaction introduces gauge
parameter dependences \cite{Baacke:1999sc,Heitmann:2001qt}, which are poorely 
understood in the framework of nonequilibrium quantum field theory. 
There is some recent progress in this direction
\cite{Arrizabalaga:2002hn,Mottola:2003vx,Carrington:2003ut}, 
however not yet on the level of concrete simulations.

\begin{acknowledgments}

The authors take pleasure in thanking H.~de Vega, 
W.~Buchm$\ddot{\textrm{u}}$ller and S.~Michalski 
for inspiring discussions.
A.H. has enjoyed very interesting discussions 
with J.~Berges, A.~Arrizabalaga, T.~Fugleberg and M.~Salle. 
A.H. thanks the Graduiertenkolleg ``Physik der Elementarteilchen 
an Beschleunigern und im Universum'' for partial financial support.

\end{acknowledgments}


\appendix
\renewcommand{\theequation}{\Alph{section}\arabic{equation}}

\section{Identities for Feynman integrals}
\label{sec:feynman-iden}
\begin{figure}[htbp]
  \centering
(a) \includegraphics[scale=0.4]{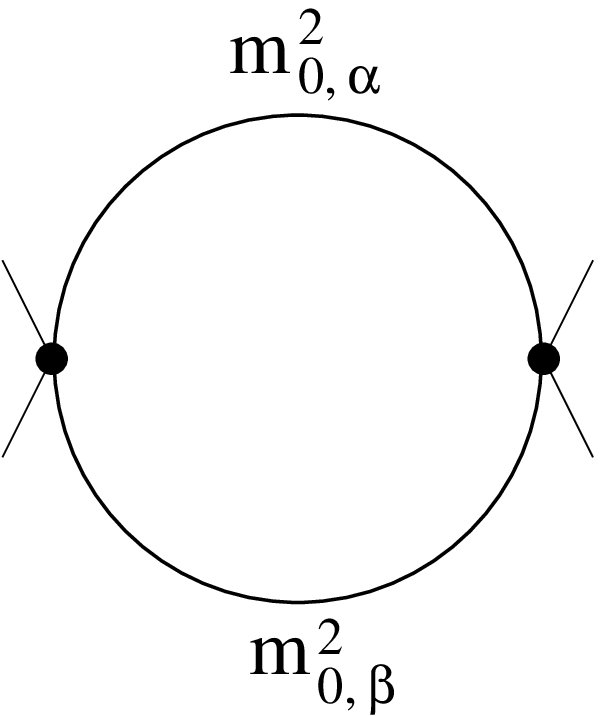}  \hspace{1cm}
(b) \includegraphics[scale=0.4]{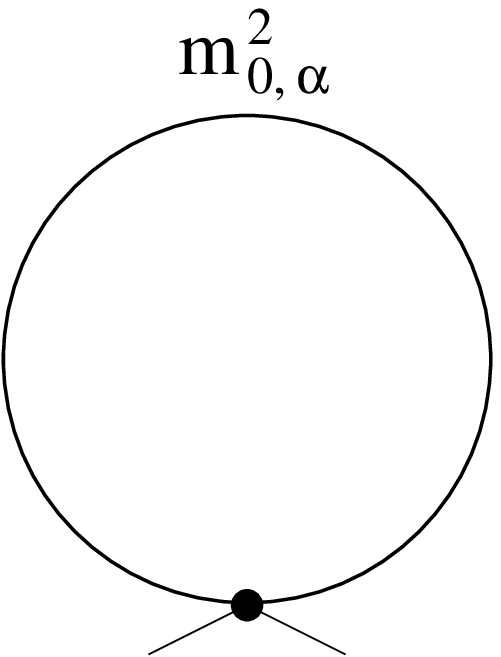}
  \caption{(a) The Feynman diagram 
with a topology of a fish graph corresponding to Eq.~(\ref{eq:fi-fish}) 
(b) the tadpole type graph corresponding to Eq.~(\ref{eq:fi-tadpole}); 
in both diagrams the lines denote free propagators 
with the initial masses $m_{0,\,\alpha}^2$ and $m_{0,\,\beta}^2$, 
respectively}
  \label{fig:fi-graphs}
\end{figure}

Within dimensional regularization ($D=4-\epsilon$) 
the following identities (no
summation over Greek indices) hold
\bea
&&\int\frac{d^{D-1}p}{(2\pi)^{D-1}}\frac{1}{2\omega_\alpha
  \omega_\beta(\omega_\alpha+\omega_\beta)}\nonumber \\
&&=\int\frac{d^{D}p}{(2\pi)^{D}}\frac{1}{(p^2-m_{0,\,\alpha}^2+io)
(p^2-m_{0,\,\beta}^2+io)}\label{eq:fi-fish} \\ 
&&=\frac{1}{16\pi^2}\left[L_\epsilon
-\ln\frac{m_{0,\,\alpha}^2}{\mu^2}+1
+\frac{m_{0,\,\beta}^2}{m_{0,\,\alpha}^2
-m_{0,\,\beta}^2}\ln\frac{m_{0,\,\beta}^2}{m_{0,\,\alpha}^2}\right]
\nonumber \\
&&
\eea
and 
\bea
\int\frac{d^{D-1}p}{(2\pi)^{D-1}}\frac{1}{2\omega_\alpha}&=&\int\frac{d^Dp}{(2\pi)^D}\frac{i}{p^2-m_{0,\,\alpha}^2+io}\\
&=&\,-\frac{m_{0,\,
    \alpha}^2}{16\pi^2}
\left[L_\epsilon-\ln\frac{m_{0,\,\alpha}^2}{\mu^2}+1\right]\quad
\label{eq:fi-tadpole} 
\eea
with
\bea
\omega_\alpha&=&\sqrt{m_{0,\, \alpha}^2+\vec{p}^2}\\
L_\epsilon&=&\frac{2}{\epsilon}-\gamma+\ln 4\pi \ .
\eea
The corresponding Feynman diagrams are depicted 
in Fig.~\ref{fig:fi-graphs}.
Note that
\bea
\lim_{m_{0,\,\beta}^2\to\,
  m_{0,\,\alpha}^2}\!\!\!&&\!\!\!
\left[-\ln\frac{m_{0,\,\alpha}^2}{\mu^2}+1
+\frac{m_{0,\,\beta}^2}{m_{0,\,\alpha}^2
-m_{0,\,\beta}^2}\ln\frac{m_{0,\,\beta}^2}{m_{0,\,\alpha}^2}\right]
\tcsplit
=
-\ln\frac{m_{0,\,\alpha}^2}{\mu^2} \ .
\eea

An identity that is needed for the renormalization 
of the energy is given by
\bea
\int\frac{d^{D-1}p}{(2\pi)^{D-1}}\omega_\alpha
=\,\frac{m_{0,\,\alpha}^4}{32\pi^2}
\left[L_\epsilon-\ln\frac{m_{0,\,\alpha}^2}{\mu^2}+\frac{3}{2}\right]
\label{eq:quartic-div} \ .
\eea


\section{Perturbative expansion of the mode functions}
\label{sec:pert-expans}
We will present in this section the isolation 
of the divergences via a perturbative expansion 
of the mode functions (see e.g. Ref.~\cite{Baacke:1997se,Baacke:1997kj}) 
for the case of a coupled system of equations. 

Let us split the mode functions $f_i^\alpha$ 
into a free part containing the initial matrix $O_{ij}$
and higher order terms represented by the 
reduced mode functions $h_i^\alpha$, i.e.
\bea
f_i^\alpha(t;p)&=&e^{-i\omega_\alpha t}
\left[O_{i\alpha}+h_i^\alpha(t;p)\right] 
\label{eq:fialpha-expans}\ .
\eea
Here and in the following no summation 
over Greek indices is meant if not explicitly stated.

If we define a potential 
\bea
\mathcal{V}_{ij}(t)&=&\calm^2_{ij}(t)-\calm^2_{ij}(0)\ 
\eea
the differential equation~(\ref{eq:mode-equations}) 
is equivalent to the following integral equation
\bea
f_{i}^\alpha(t;p)&=&e^{-i\omega_\alpha
  t} O_{i\alpha} 
\tcsplit
+\int_0^t{dt'\,} K^\mathrm{ret}_{ij}(t-t';p)
\mathcal{V}_{jk}(t')f_k^\alpha (t';p)\nonumber \ .
\eea
The retarded kernel of the free equation is given by
\bea
K^\mathrm{ret}_{ij}(t-t';p)&=&
\sum_{\beta}\frac{i}{2\omega_\beta}\Theta(t-t')O_{i\beta}O_{j\beta}
\tcsplit
\times\left[e^{i\omega_\beta(t-t')}
-e^{-i\omega_\beta(t-t')}\right] \ .
\eea 

Inserting the retarded kernel in the integral equation gives
\bea 
f_i^\alpha(t;p)&=&e^{-i\omega_\alpha
  t}O_{i\alpha}
\tcsplit
+\int_0^t{dt'\,}\sum_\beta
\frac{i}{2\omega_\beta}O_{i\beta}O_{j \beta}
\tcsplit
\times
\left[e^{i\omega_\beta(t-t')}
-e^{-i\omega_\beta(t-t')}\right]
\tcsplit
\times 
\mathcal{V}_{jk}(t') f_{k}^\alpha(t';p)
\label{eq:integral-eq-f}\\
&=&e^{-i\omega_\alpha
  t}\bigg\{O_{i\alpha}+\int_0^t{dt'\,}\mathcal{V}_{jk}(t') 
\label{eq:inteqf}
\\
&&
\qquad\quad\times
\sum_\beta
\frac{i}{2\omega_\beta}O_{i\beta}O_{j\beta}O_{k\alpha}
\tcsplit
\quad \times
\bigg[e^{i(\omega_\beta+\omega_\alpha)
  (t-t')}
\tcsplit
\qquad \quad 
-e^{-i(\omega_\beta -\omega_\alpha)(t-t')}\bigg]
\bigg\}\nonumber \\
&& +\ldots\ , 
\eea 
where in addition the decomposition in Eq.~(\ref{eq:fialpha-expans}) 
has been used. The dots imply the higher order terms with $h_i^\alpha$ 
that we do not need for the analysis of the divergences here.

By partial integration the divergent
contributions can be isolated in the usual manner 
\bea
f_i^\alpha(t;\vec{p})&=&e^{-i\omega_\alpha
  t}\bigg\{O_{i\alpha}
\label{eq:inteqf-2} 
\\
&&\quad
-\sum_\beta
\frac{1}{2\omega_\beta}\left[\frac{1}{\omega_\beta+\omega_\alpha}
+\frac{1}{\omega_\beta-\omega_\alpha}\right]
\tcsplit
\quad \times
O_{i\beta}O_{j\beta}O_{k\alpha}\mathcal{V}_{jk}(t)
\nonumber \\
&&\quad +\sum_\beta\frac{1}{2\omega_\beta}
\bigg[\frac{1}{\omega_\beta+\omega_\alpha}e^{i(\omega_\beta+\omega_\alpha)t}
\tcsplit
\quad
+\frac{1}{\omega_\beta-\omega_\alpha}
e^{-i(\omega_\beta-\omega_\alpha)t}\bigg]
\tcsplit
\quad\times 
O_{i\beta}O_{j\beta}O_{k\alpha}\mathcal{V}_{jk}(0)\nonumber \\
&&\quad +\sum_\beta
\frac{1}{2\omega_\beta}
O_{i\beta}O_{j\beta}O_{k\alpha}
\tcsplit
\quad\times
\int_0^tdt'\dot{\mathcal{V}}_{jk}(t')
\bigg[\frac{1}{\omega_\beta+\omega_\alpha}
e^{i(\omega_\beta+\omega_\alpha)(t-t')}\nonumber \\
&&\qquad +\frac{1}{\omega_\beta-\omega_\alpha}
e^{-i(\omega_\beta-\omega_\alpha)(t-t')}\bigg]
\bigg\}\nonumber\\ 
&&+\ldots \nonumber\ .
\eea
This expression and its complex conjugate is all what is needed to 
calculate the divergent contributions in $\Delta_{ij}(t)$. 
We have for the divergent part of the symmetrized Green's function
\bea
&&\frac{1}{2}\bigg[G_{ij}(t,t;\vec{p})+G_{ji}(t,t;\vec{p})\bigg]^\mathrm{div}
\tcsplit
=\sum_\alpha
\frac{1}{2\omega_\alpha}\mathrm{Re}
\bigg[f_i^\alpha(t;p)f_j^{\alpha,\,*}(t;p)
\bigg]^\mathrm{div}\nonumber \\
&&=\sum_\alpha
\frac{1}{2\omega_\alpha}\bigg[O_{i\alpha}O_{j\alpha}
\tcsplit
\qquad 
+\sum_\beta
\frac{1}{2\omega_\beta}
O_{i\beta}O_{l\beta}O_{k\alpha}O_{j\alpha}\mathcal{V}_{kl}(t)
\frac{-2\omega_\beta}{\omega_\beta^2-\omega_\alpha^2}\nonumber
\\
&&\qquad +\sum_\beta
\frac{1}{2\omega_\beta} O_{i\alpha}O_{l\beta}O_{k\alpha}O_{j\beta}
\mathcal{V}_{kl}(t)
\frac{-2\omega_\beta}{\omega_\beta^2-\omega_\alpha^2}\bigg]\nonumber \\
&&=\sum_\alpha
\frac{1}{2\omega_\alpha}\bigg[O_{i\alpha}O_{j\alpha}
\tcsplit
\qquad
-\frac{1}{\omega_\beta(\omega_\alpha+\omega_\beta)}
O_{i\alpha}O_{j\beta}O_{l\beta}O_{k\alpha}\mathcal{V}_{kl}(t)\bigg]
\label{eq:Delta-int-div}\ .
\eea
These expressions become divergent if they are integrated over $d^3p$. 
We will use them as subtraction terms in the fluctuation integrals in 
$\Delta^{(1)}_{ij}(t)$ and $E^{(1)}(t)$ [see Eqs.~\eqn{eq:Delta_ij_1} 
and \eqn{eq:energy-E1}].

The divergences in the energy can be found in an analogous
way by inserting $f_i^\alpha(t,p)$ and its time derivative 
in the one-loop energy part $E^{(1)}(t)$ [see Eq.~\eqn{eq:energy-E1}].
We will use an alternative approach in the following.

The one-loop effective action at time $t$ minus the one at
$t=0$ is given by
\be
\widetilde{\Gamma}^{(1)}[\calm^2]=\frac{i}{2}\tr \ln \left\{
\frac{\Box  + \calm^2(t)}{\Box  +  \calm^2(0)}
\right\}
\ee
where it is understood that the numerator and denominator are
$2\times2$ matrices.
This expression can be expanded locally with respect to
$\calv=\calm^2(t)-\calm^2(0)$ and gradients thereof.
The expansion can be obtained by going to the momentum 
representation and by expanding with respect to insertions
of $\calv(q)$ and with respect to the external momenta 
$q=(q_0,\mathbf{q})$. As we do not need an
infinite wave function renormalization the divergent parts
are given by the terms of first and second order in $\calv(q)$.

We introduce
\bea
G_{0,ij}^{-1}(p)&=&\left(-p_0^2+\bfp^2\right)\delta_{ij}
+m^2_{0,ij}
\\
G_{ij}^{-1}(p)&=&\left(-p_0^2+\bfp^2\right)\delta_{ij}
+\calm^2_{ij}(t)
\eea
with $m^2_{0,ij}=\calm^2_{ij}(0)$. $G_0$ is  {\em not } the 
bare propagator which would be defined
at the vacuum expectation values of $\phi$ and $\chi$.
We diagonalize the initial mass matrix by an orthogonal transformation
\be
 \calm^2(0)= O \widetilde\calm^2(0) O^T
\ee
or 
\be
 m^2_{0,ij}=O_{i\alpha}O_{j\beta} \widetilde m^2_{0,\alpha \beta}
=O_{i\alpha}O_{j\alpha} \widetilde m^2_{0,\alpha}
\pkt\ee
Then also $G_{0,ij}^{-1}$ becomes diagonal. 
The same holds true for the inverse matrix.
We likewise introduce
\bea
G_{ij}^{-1}(p)&=&O_{i\alpha}
\left(\left[-p_0^2+\bfp^2+\widetilde m^2_{0,\alpha}\right]\delta_{\alpha\beta}
+\widetilde \calv_{\alpha\beta}(t)\right)O_{j\beta}\nonumber \\
&=&O_{i\alpha}\widetilde G_{\alpha\beta}^{-1}(p)O_{j\beta} \ ,
\eea
where of course 
\be
\widetilde \calv_{\alpha\beta}(t)=
O_{i\alpha}\calv_{ij}(t)O_{j\beta}\label{eq:redef_vtilde}
\ee is no longer diagonal.
The effective action, in the approximation where all gradient terms
are neglected, can now be rewritten as
\bea \nonumber
 \widetilde{\Gamma}^{(1)}&\simeq&
\frac{i}{2}\int\frac{d^4p}{(2\pi)^4}
\tr\ln\left\{G_0 G^{-1}\right\}
\\&=&\int\frac{d^4p}{(2\pi)^4}
\tr\ln\left\{1+\widetilde G_0 \widetilde\calv(t)\right\}
\pkt\eea
The first terms in the expansion are
\bea
 \widetilde{\Gamma^{(1)}}&\simeq&
\frac{i}{2}\int\frac{d^4p}{(2\pi)^4}
\left[\sum_\alpha\frac{1}{-p_0^2+\bfp^2+m^2_{0,\alpha}+io}
\widetilde\calv_{\alpha\alpha}(t)\right.
\nonumber\\
&&-\frac{1}{2} \sum_{\alpha\beta}
\frac{1}{-p_0^2+\bfp^2+m^2_{0,\alpha}+io}
\widetilde\calv_{\alpha\beta}(t)
\tcsplit
\times
\left. \frac{1}{-p_0^2+\bfp^2+m^2_{0,\beta}+io}
\widetilde\calv_{\beta\alpha}(t)
\right]\nonumber\\
&& 
+ O(\widetilde \calv^3)
\eea
The three-dimensional reduction is obtained via 
Eq.~\eqn{eq:fi-tadpole} and Eq.~\eqn{eq:fi-fish}.
So we find
\bea
\widetilde{\Gamma}^{(1)\, {\rm div}}&=&\int\frac{d^3p}{(2\pi)^3}
\bigg[\sum_\alpha\frac{-\widetilde\calv_{\alpha\alpha}(t)}{4\omega_\alpha}
\tcsplit
-\frac{1}{2}\sum_{\alpha\beta}
\frac{-\widetilde\calv_{\alpha\beta}(t)
\widetilde\calv_{\beta\alpha}(t)}{4\omega_\alpha\omega_\beta
(\omega_\alpha+\omega_\beta)}
\bigg] \ .
\eea
The divergent parts of the fluctuation energy are, therefore,
\be
E^{(1)}_{\rm div}=\int\frac{d^3p}{(2\pi)^3}
\left[\sum_\alpha\frac{\widetilde\calv_{\alpha\alpha}(t)}{4\omega_\alpha}
-\sum_{\alpha\beta}
\frac{\widetilde\calv_{\alpha\beta}(t)
\widetilde\calv_{\beta\alpha}(t)}{8\omega_\alpha\omega_\beta
(\omega_\alpha+\omega_\beta)}
\right] \ .\ \label{eq:E1_div}
\ee
As a cross check we may obtain the divergent terms in the fluctuation
integrals $\Delta_{ij}$ which are given by
\be
\frac{\Delta^{(1)\, \rm div}_{ij}}{2}=-\frac{\delta \widetilde{\Gamma}^{(1)\, \rm div}}
{\delta \calm^2_{ij}(t)}=-
\frac{\delta \widetilde{\Gamma}^{(1)\, \rm div}}{\delta \calv_{ij}(t)}
\ee
Using Eq.~\eqn{eq:redef_vtilde} we have
\be
\frac{\delta \calv_{\alpha\beta}(t)}{\delta \calv_{ij}(t)}
=O_{i\alpha}O_{j\beta}
\ee 
and therefore
\bea
\Delta^{(1)\,\rm div}_{ij}&=&
\int\frac{d^3p}{(2\pi)^3}
\bigg[\sum_\alpha\frac{1}{2\omega_\alpha}O_{i\alpha}O_{j\alpha}
\tcsplit
-\sum_{\alpha\beta}
\frac{O_{i\alpha}O_{j\beta}
\widetilde\calv_{\beta\alpha}(t)}{2\omega_\alpha\omega_\beta
(\omega_\alpha+\omega_\beta)} 
\bigg]\ .
\eea


\section{Counterterm structure}
\label{sec:2PPI}
In this appendix we will establish a connection between the effective
counterterms used above and the counterterms in the 2PPI formalism. 
The latter ones follow from a standard counterterm Lagrangian. 
For details we refer the reader to 
Ref.~\cite{Verschelde:1992bs:Coppens:1993zc,Verschelde:2000ta,
Verschelde:2000dz}.
 
The general Lagrange density for the model in Eq.~\eqn{eq:general-lag} 
including a counterterm $\delta \mathcal{L}$  denotes
\bea
\call&=&\frac{1}{2}\del_\mu\Phi^i\del^\mu\Phi^i
-\frac{1}{2}m_{ij}^2\Phi^i\Phi^j\nonumber
-\frac{1}{4!}\lambda_{ijkl}\Phi^i\Phi^j\Phi^k\Phi^l\\
&&+\delta
\call\ . 
\eea
The counterterm Lagrangian $\delta \mathcal{L}$ is given by
\bea
\delta \call&=&\frac{1}{2}\delta
Z_{ij}\del^\mu\Phi^i\del^\mu\Phi^j-\frac{1}{2}\delta m_{ij}^2
\Phi^i\Phi^j
\nonumber \\
&&-\frac{1}{4!}\delta\lambda_{ijkl}\Phi^i\Phi^j\Phi^k\Phi^l 
\label{eq:delta-cal_L}\ .
\eea
The gap equation for the renormalized effective mass 
$\calm^2_{\mathrm{R},\,ij}$ with mass and coupling constant counterterms
 is given by
\bea
\calm^2_{\mathrm{R},\,ij}&=&m_{ij}^2+\delta
m_{ij}^2\nonumber \\
&&+\frac{1}{2}(\lambda_{ijkl}+\delta\lambda_{ijkl})
\left(\phi^k\phi^l+\Delta^{kl}\right) \ .
\eea
In the 2PPI scheme the following relations for the renormalization constants hold, 
if one takes a \emph{mass independent renormalization scheme} 
\bea
(\delta m^2)^{ij}&=&\delta Z_{m}^{ij;kl}m_{kl}^2\\
\delta \lambda^{ij;kl}&=&\lambda^{ij}_{\ \ pq}\delta Z_{m}^{\, pq;kl}\\
\delta Z_{m}^{\, ij;kl}&=&\lambda^{ij}_{\ \ pq}\delta \zeta^{pq;kl}\ .
\eea
Thus all renormalization constants can be derived from a  
vacuum counter term
\bea
\delta E_\mathrm{vac}&=&\frac{1}{2}\calm^2_{ij}\calm^2_{kl}\,\delta
\zeta^{ij;kl} \label{eq:E_vac}\ .
\eea

With the help of these identities the gap equation reads
\bea
\calm^2_{\mathrm{R},\,ij}&=&m_{ij}^2
+\frac{1}{2}\lambda_{ijkl}\left(\phi^k\phi^l+\Delta^{kl}\right)\nonumber
\\
&&+\lambda_{ijpq}\,\delta \zeta^{pq;kl}m_{kl}^2\nonumber
\\
&&+\frac{1}{2}\lambda_{ijpq}\lambda^{pqrs}\,\delta \zeta_{rs;kl}
\left(\phi^k\phi^l+\Delta^{kl}\right)\\
&=&m_{ij}^2
+\frac{1}{2}\lambda_{ijkl}\left(\phi^k\phi^l+\Delta^{kl}_\mathrm{R}\right)
\ ,
\eea
with
\bea
\lambda_{ijkl}\Delta^{kl}_\mathrm{R}&=&\lambda_{ijkl}\Delta^{kl}
+2\lambda_{ijpq}\delta \zeta^{pq;kl}m_{kl}^2\nonumber
\\
&&+\frac{1}{2}\lambda_{ijpq}\lambda^{pqrs}\,\delta \zeta_{rs;kl}
\left(\phi^k\phi^l+\Delta^{kl}\right) \label{eq:Delta_ij_R}\ .
\eea
Once the propagator $\Delta^{ij}$ is renormalized the gap equations 
for the masses $\calm^2_{ij}$ become finite. In the Hartree 
approximation, i.e. the one-loop 2PPI approximation, the equations 
for the classical fields $\phi^i$ are finite as well. 
No wave function renormalization is needed.

If we specialize Eq.~(\ref{eq:Delta_ij_R})  
with the help of Eq.~(\ref{eq:id1})--(\ref{eq:id8}) 
to our given hybrid model, 
the renormalization constants $\delta \zeta$ simplify to  
\bea
\delta\zeta_{\phi\phi;\phi\phi}&=&\delta\zeta_{\chi\chi;\chi\chi}
=:-2\delta\xi_{\phi\phi}=-2\delta\xi_{\chi\chi}\\
\delta\zeta_{\phi\chi;\phi\chi}&=&\delta\zeta_{\chi\phi;\chi\phi}
=:-2\delta\xi_{\phi\chi}\\
\delta\zeta_{\phi\phi;\chi\chi}&=&\delta\zeta_{\phi\chi;\phi\phi}
=\ldots=0\ .
\eea
The vacuum counter term in Eq.~(\ref{eq:E_vac}) reduces to 
\bea
\delta E_\mathrm{vac}&=&-\delta\xi_{\phi\phi}(\calm^2_{\phi\phi})^2
-\delta\xi_{\chi\chi}(\calm^2_{\chi\chi})^2
-2\delta\xi_{\phi\chi}(\calm^2_{\phi\chi})^2\nonumber \\
&=&-\delta \calm^4 \ .
\eea
Thus the effective counterterm $\delta \calm^4$ used in 
Sec.~\ref{subsec:effect} has been mapped to 
the renormalization constants of a standard counterterm Lagrangian 
$\delta \mathcal{L}$ [see Eq.~(\ref{eq:delta-cal_L})]. In principle
one can calculate the nonperturbatively fixed 
counterterms $\delta m^2$, $\delta v^2$, 
$\delta g^2$ and $\delta \lambda$ explicitly, 
insofar the renormalized gap equations form a coupled system of equations. 
However, this would not be very enlightening.


\end{document}